\documentclass[conference]{IEEEtran}
\usepackage{booktabs}
\usepackage{graphicx}
\usepackage{subcaption}
\usepackage{amsmath}

\usepackage{amssymb}
\usepackage{xspace}
\usepackage{pifont}
\usepackage{threeparttable}
\usepackage{multirow}
\usepackage{enumitem}
\usepackage{algorithm}
\usepackage{algpseudocode}
\usepackage[table]{xcolor}
\usepackage{wrapfig}
\usepackage{makecell}
\usepackage{url}
\usepackage{hyperref}
\usepackage[nameinlink,capitalize]{cleveref}
\usepackage[numbers,sort&compress]{natbib}

\newcommand{\bx}{\mathbf{x}}
\newcommand{\Wass}{W_2}

\newtheorem{theorem}{Theorem}

\definecolor{randcol}{RGB}{214,183,165}
\definecolor{kindonlycol}{RGB}{190,143,112}
\definecolor{kindsemcol}{RGB}{123,125,103}
\definecolor{fullfpcol}{RGB}{165,162,132}
\definecolor{theorycol}{RGB}{205,203,188}

\begin{document}

\title{MDRC: A Deployable State-Recovery Defense for Traffic Signal Control under Sensor Corruption}
\author{\IEEEauthorblockN{Mingyuan Li\IEEEauthorrefmark{1}, Chunyu Liu\IEEEauthorrefmark{2}, Xiao Liu\IEEEauthorrefmark{2}, Yanna Jiang\IEEEauthorrefmark{3}, Guangsheng Yu\IEEEauthorrefmark{4}, Xu Wang\IEEEauthorrefmark{4}, Wei Ni\IEEEauthorrefmark{5}, Ren Ping Liu\IEEEauthorrefmark{4}}\IEEEauthorblockA{\IEEEauthorrefmark{1}ELLIS Institute Finland, University of Turku, Turku, Finland\\\IEEEauthorrefmark{2}Independent Researcher\\\IEEEauthorrefmark{3}City University of Macau\\\IEEEauthorrefmark{4}University of Technology Sydney\\\IEEEauthorrefmark{5}Edith Cowan University}}

\IEEEoverridecommandlockouts
\makeatletter\def\@IEEEpubidpullup{6.5\baselineskip}\makeatother
\IEEEpubid{\parbox{\columnwidth}{
		Network and Distributed System Security (NDSS) Symposium 2027\\
		22--26 March 2027, Seoul, Republic of Korea\\
		ISBN 978-1-970672-09-1\\  
		https://dx.doi.org/10.14722/ndss.2027.[23$|$24]xxxx\\
		www.ndss-symposium.org
}
\hspace{\columnsep}\makebox[\columnwidth]{}}

\maketitle

\begin{abstract}
Traffic Signal Control (TSC) is a safety-critical cyber-physical system that relies on real-time sensing. In practice, corrupted observations caused by adversarial perturbations or sensor failures can directly propagate from the sensing layer into the controller and degrade traffic efficiency. Existing robust Reinforcement Learning(RL)-based TSC methods improve performance in controlled settings, but suffer from limited cross-city generalization, high inference latency, and weak recovery under partial observability.
In this paper, we present MDRC (\textbf{M}eta-\textbf{D}iffusion-based framework for \textbf{R}esilient traffic signal \textbf{C}ontrol against adversarial attacks and sensor failures), a post-detection state recovery defense inserted between sensing and control. MDRC reconstructs trustworthy traffic states from corrupted observations before they are consumed by the downstream controller, thereby decoupling control decisions from attacked inputs. We combine Denoising Diffusion Implicit Models (DDIM)-based state recovery with Reptile meta-learning:
Reptile learns a transferable recovery-model initialization across
cities, while DDIM provides efficient inference-time reconstruction
of corrupted states. We provide an optimization-based view of
the DDIM recovery dynamics and establish a recovery-error bound that
separates score approximation, numerical discretization, and
initialization mismatch.
Across seven real-world-derived CityFlow benchmarks, MDRC reduces Average Travel Time(ATT) by 6.77\% on average under stochastic and policy-aware attacks and by 12.75\% under structured sensor loss, while improving state-recovery fidelity. Beyond simulation, we conduct a controlled sensor-failure experiment on 3,600 seconds of real roadside measurements, disabling 50\% of detector channels and evaluating whether MDRC can recover the unavailable sensing information before control. We further integrate MDRC into a hardware-in-the-loop traffic-signal stack containing physical roadside sensors, edge acquisition, GPU inference, and a commercial signal controller. Over a 9.16-hour run comprising 32,389 sensing/control cycles, the system achieves 99.79\% decision
availability, produces no out-of-plan recommendations, and requires
approximately 38~ms of component-wise processing per one-second
control interval. These results establish prototype-level real-data
recovery and real-time system-integration feasibility.

\end{abstract}

\section{Introduction}

Traffic Signal Control (TSC) is a fundamental component of urban traffic management and a representative cyber-physical system whose decisions depend on continuous sensing of the physical environment~\citep{liang2018deep}. Modern TSC pipelines rely on cameras, loop detectors, or radars to estimate queue lengths, occupancies, and vehicle movements, then feed these observations into rule-based or learning-based controllers~\citep{CoLight2019}. This sensing-to-control path is a critical attack surface: noisy measurements, sensor outages, or adversarial perturbations at the observation layer can be directly translated into poor signal-phase decisions, amplifying congestion and degrading system reliability.
Recent robust Reinforcement Learning(RL) methods, especially RobustLight~\citep{lirobustlight,li2025fuzzylightrobusttwostagefuzzy}, mark an important step toward resilient TSC. However, several limitations prevent them from meeting the needs of security-critical deployment:
1) its per-city training paradigm requires a separate model for each environment, restricting cross-city deployment and transfer;
2) its inference and adaptation overhead is high, making it difficult to insert into real-time traffic control loops; and
3) it is still best viewed as a robust-learning method, rather than a security mechanism that explicitly defends the sensing-to-control interface under adversarial and sensor-failure conditions.

From a security perspective, the core problem is not simply to learn a better controller, but to recover a trustworthy traffic state once sensing becomes corrupted. In practical deployments, anomaly detectors, sensor-health monitors, or rule-based consistency checks often exist upstream. Once such a mechanism flags suspicious sensing, the next question is how to mitigate the corruption before the controller acts on it. This post-detection recovery problem is largely independent of attack detection itself, yet it is the step that determines whether the system can continue operating safely and efficiently.
To address this challenge, we introduce \textbf{MDRC}, a post-detection \emph{state recovery defense mechanism} that is inserted between sensing and control. Rather than retraining or replacing the downstream controller, MDRC reconstructs a clean state estimate from the compromised observation and forwards the recovered state to the original policy. This design decouples control from corrupted input, and because it operates on the state interface, it is compatible with both RL-based and rule-based controllers.

Two technical ingredients make this defense practical. First, to achieve cross-city transfer, we meta-train the recovery model across heterogeneous cities so that it can be deployed zero-shot and adapted few-shot in unseen environments~\citep{huang2021modellight,MetaLight}. Second, we build the recovery mechanism on Denoising Diffusion Implicit Models (DDIM)~\citep{DDIM}, whose fast sampling substantially reduces inference latency compared with prior diffusion-based defenses.
We combine Reptile-based parameter-space meta-learning with DDIM-based state-space recovery. Reptile~\citep{reptile} learns a transferable recovery-model initialization across source cities, while DDIM provides efficient inference-time reconstruction of corrupted observations. We further provide an optimization-based interpretation of the DDIM recovery dynamics and analyze the resulting recovery error.
Based on this connection, MDRC unifies diffusion-based recovery and meta-learning, enabling rapid adaptation across cities with only a few fine-tuning steps while remaining fast enough for online deployment.

This security-centric formulation changes the role of diffusion from an ML robustness trick into a deployable defense layer for resilient TSC.
Our primary contributions are:
\begin{itemize}
    \item We formalize resilient TSC as a post-detection state recovery problem on the sensing-to-control path, and define a concrete observation-space threat model covering stochastic attacks, adaptive policy-targeted attacks, and physical sensor failures.
    \item We provide an optimization-based interpretation of DDIM recovery dynamics and clarify how inference-time state-space denoising complements Reptile-based parameter-space meta-learning for cross-city recovery.
    \item We propose \textbf{MDRC}, a plug-in, controller-agnostic defense mechanism that reconstructs clean traffic states before control, enabling superior cross-city generalization and rapid adaptation with a single shared model.
    \item We validate MDRC on seven real-world-derived benchmarks, a measured roadside-detector trace, and a hardware-in-the-loop testbed with physical sensors and a commercial signal controller, demonstrating robust transfer and real-time deployment feasibility.
\end{itemize}

\section{Related Works}
TSC has progressed from rule-based methods ~\citep{webster1958} to adaptive systems like SCOOT ~\citep{hunt1981scoot}, SCATS ~\citep{SCATS}, and RHODES ~\citep{MIRCHANDANI2001415}, which reduce delays dynamically. Control-theoretic methods such as Max Pressure~\citep{varaiya2013max} and Efficient Pressure~\citep{wu2021efficient} offer robustness but rely on simplified assumptions. RL has become the dominant paradigm~\citep{wei2021recent}, with approaches like IntelliLight~\citep{wei2018intellilight}, PressLight~\citep{wei2019presslight}, CoLight, and MetaLight~\citep{MetaLight} achieving strong results.
Recent works explore multi-agent coordination~\citep{Cooperative2024} and multi-modal or hierarchical representations~\citep{Decentralized2023, large2024, ruan2024coslight, duan2025bayesian}, yet most still rely on fixed, hand-crafted state features. Viewing traffic as a complex system~\citep{mitchell2009complexity, strogatz2001nonlinear}, hierarchical RL provides scalable solutions. 

Meta-learning has advanced through diverse strategies, including latent embedding optimization ~\citep{rusu2018meta}, differentiable convex solvers ~\citep{lee2019meta}, implicit gradients ~\citep{rajeswaran2019meta, zhang2023scalable}, and sparsity-aware adaptation ~\citep{von2021learning}. Studies have also explored the trade-off between rapid learning and feature reuse ~\citep{raghu2019rapid}. Recently, MetaDiff ~\citep{zhang2024metadiff} introduced a task-conditional diffusion-based framework that generalizes gradient descent with learnable momentum and uncertainty modeling. Meta-learning has recently emerged as a promising direction in TSC for improving generalization and adaptability across varying traffic scenarios. Early works like MetaLight ~\citep{MetaLight} and CrossLight ~\citep{sun2024crosslight} applied meta-learning for quick adaptation and cross-scenario generalization, but overlooked safety concerns in TSC.

Diffusion models have achieved remarkable success in generative modeling ~\citep{beta_schedule3, DDIM, nichol2021improved}, with improvements in training stability and sample quality through advanced beta schedules ~\citep{beta_schedule1}. Classifier-free guidance enhanced controllability ~\citep{ho2022classifier}. Recently, diffusion has been extended to RL for robust decision-making under uncertainty, such as DiffLight for missing data in TSC ~\citep{chen2024difflight}, RobustLight for policy robustness ~\citep{lirobustlight}, and DMBP for offline RL with noisy states ~\citep{yang2023dmbp}. These algorithms suffer from slow inference speed, making them impractical for real-world deployment, and their reconstruction performance remains suboptimal.

\textbf{Recovery in cyber-physical control systems.}
Recovery has been studied for UAVs and robotic autonomous vehicles,
where existing methods typically recover the state or control actions
of a single mobile platform~\cite{fei2020learn,dash2024specguard}.
In TSC, falsified connected-vehicle data can manipulate signal decisions
and cause traffic congestion~\cite{chen2018exposing,
huang2021impact}, while traffic-data anomaly detection provides a
complementary upstream capability~\cite{sarteshnizi2023traffic}.
In contrast, MDRC focuses on post-detection recovery in a networked
infrastructure system by reconstructing movement-level traffic states
before an unchanged controller. Compared with classical model-based
estimators such as Kalman filtering~\cite{kalman1960new}, MDRC targets
nonlinear cross-city distributions, heterogeneous corruption, and
structured channel loss without requiring a separately calibrated
dynamics model for each city.

\section{Preliminary}

\noindent\textbf{Cross-City Few-Shot State-Recovery Transfer.} We consider a meta-learning setting for traffic-state recovery. Source cities provide offline sensing and control trajectories, while
a target city provides either no calibration data for zero-shot deployment or a small number of measured state sequences for few-shot adaptation. The objective is to transfer a shared recovery model that reconstructs trustworthy traffic states under corrupted or missing observations. MDRC does not transfer, retrain, or replace the
downstream TSC policy.

\begin{figure}[t] 
    \centering
    \includegraphics[width=1.0\linewidth]{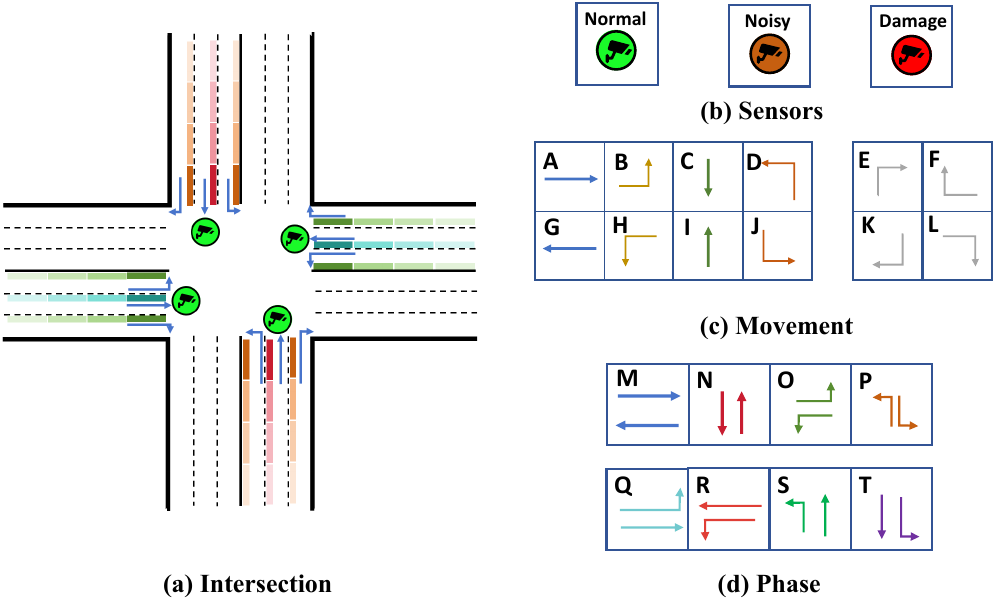} 
    \caption{Definition of the TSC. Sensors are used to acquire vehicle information, which is fed into the TSC algorithm to output the appropriate phase, as shown in Figure 1(d).}
    \label{fig:definition}
\end{figure}

\noindent\textbf{Traffic Signal Control.} We use a four-way intersection, as depicted in Figure~\ref{fig:definition}(a), to introduce key concepts and definitions for TSC. 
A road network comprises multiple intersections, each with $N$ road segments, denoted as $\{Inter_1, \dots, Inter_N\}$. 
Each intersection is equipped with four directional sensors (e.g., cameras, radars) monitoring three lanes per direction. 
Sensor states are color-coded as green for normal, orange for noise attacks, and red for sensor damage, as shown in Figure~\ref{fig:definition}(b). 
A vehicle's path through an intersection, from an entry lane ($lane_{in}$) to an exit lane ($lane_{out}$), is defined as $TM = (lane_{in}, lane_{out})$, as illustrated in Figure~\ref{fig:definition}(c). 
A traffic signal phase consists of two distinct movements, $TM_i$ and $TM_j$ ($i \neq j$), denoted as $p_w = (TM_i, TM_j)$, as shown in Figure~\ref{fig:definition}(d).

\noindent\textbf{Denoising Diffusion Implicit Models.} Diffusion models generate data by reversing a Markovian noise process over $T$ steps. To accelerate inference, DDIM~\citep{DDIM} introduces a strided schedule ${\tau_1, \dots, \tau_S}$ with $S \ll T$, skipping redundant steps. The transition distribution is reformulated as:
\begin{align}
\mathbf{x}_{t-1} &= \sqrt{\bar{\alpha}_{t-1}} \left( \frac{\mathbf{x}_t - \sqrt{1 - \bar{\alpha}_t} \boldsymbol{\epsilon}_\theta^{(t)}(\mathbf{x}_t)}{\sqrt{\bar{\alpha}_t}} \right) \notag \\  
&\quad + \sqrt{1 - \bar{\alpha}_{t-1} - \sigma_t^2} \boldsymbol{\epsilon}^{(t)}_\theta(x_t)+ \sigma_t \boldsymbol{\epsilon}, \notag
\end{align}
where $\boldsymbol{\epsilon}_\theta^{(t)}$ predicts the noise at step $t$, and $\sigma_t^2 = \delta \cdot \tilde{\beta}_t$ modulates stochasticity via a tunable hyperparameter $\delta > 0$, $\tilde{\beta}_t=\sigma_t^2=\frac{1-\bar{\alpha}_{t-1}}{1-\bar{\alpha}_t} \cdot \beta_t$, $\alpha_t=1-\beta_t$ and $\bar{\alpha}_t=\prod_{i=1}^t \alpha_i$ and $\sigma \sim \mathcal{N}(0, I)$. Setting $\delta = 0$ yields a deterministic generation process. DDIM preserves the marginal distribution of DDPM but allows for significantly faster sampling. To capture the underlying data geometry,~\cite{song2020score} approximates the true score $\nabla \log p_t(\mathbf{x})$ using a parameterized estimator $\mathbf{s}_\theta(\mathbf{x}, t)$. The model parameters $\theta$ are optimized by minimizing the weighted Fisher divergence, which takes the form of a weighted Mean Squared Error (MSE):
\begin{equation}
    J(\theta; \lambda) := \frac{1}{2} \int_0^T \lambda(t) \mathbb{E}_{p_t} \left[ \| \nabla \log p_t(\mathbf{x}) - \mathbf{s}_\theta(\mathbf{x}, t) \|^2 \right] dt.
    \label{eq:sm_objective}
\end{equation}
Here, $\lambda(t) > 0$ is a time-dependent weighting function that balances the learning signal across different noise scales. \cite{diffusion_warssin} estimates the Wasserstein distance $\Wass(p_0,q_0)$ between a given data distribution $p_0$ and the
marginal of the generated samples $q_0$ obtained by the score-based model $\mathbf{s}_\theta(\mathbf{x}, t)$.

\noindent\textbf{Reptile Meta-Learning.} Reptile is a first-order gradient-based meta-learning algorithm, similar in spirit to Model-Agnostic Meta-Learning (MAML)~\citep{maml} but computationally more efficient. It iteratively samples tasks, performs $k$ steps of task-specific Stochastic Gradient Descent (SGD) from an initialization $\theta$ to obtain $\phi=\nabla\mathcal{L}_\mathcal{T}(\theta)$, and updates $\theta$ toward $\phi$:
\begin{equation}
\theta \leftarrow \theta - \mu (\theta - \phi),
\label{eq:reptile}
\end{equation}
where $\mu$ is the meta step size. This process encourages $\theta$ to lie close to the optimal manifold $\mathcal{M}_i$ of each task $\mathcal{T}_i$. Assuming that each task $\mathcal{T}_i$ has an optimal parameter manifold $\mathcal{M}_i(\phi)$, the learning objective becomes minimizing the aggregated distance $|\theta - \mathcal{M}_i(\phi)|_2^2$ across tasks. The Reptile update approximates the gradient of this objective by treating $\phi$ as a proxy for the projection of $\theta$ onto $\mathcal{M}_i(\phi)$:
\begin{equation}
\nabla_\theta [\frac{1}{2}|\theta - \mathcal{M}_i(\phi)|_2^2] \approx (\theta - \phi).
\end{equation}
This enables generalization by locating $\theta$ near the intersection of task-specific optima, facilitating fast adaptation in new tasks.

\section{Threat Model and Attack Taxonomy}
\label{section:threat_model}
In this section, we define the security assumptions used throughout the paper, focusing on \emph{post-detection recovery} against observation-space attacks on the sensing pipeline of a deployed TSC system.

\begin{figure}[t] 
    \centering
    \includegraphics[width=1.0\linewidth]{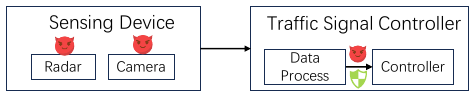} 
    \caption{Threat model description.}
    \label{threat_model}
\end{figure}

\noindent\textbf{Attack Surface.} As shown in Figure~\ref{threat_model}, the attacker targets the observation vector $\mathbf{s}_t$ produced by the sensing layer before it reaches the controller, via sensor corruption, communication disruption, or partial sensor failure. We do not consider attacks on controller logic, model parameters, rewards, training data, or actuation hardware, isolating the sensing-to-control interface targeted by MDRC.

\noindent\textbf{Attacker Capabilities and Objectives.} We consider a bounded adversary aiming to increase congestion or travel time by corrupting observations. Perturbations are constrained by an intensity $k$ or an $\ell_\infty$ ball $\mathbf{B}_d(\mathbf{s}, k)$. We study two regimes: (i) \emph{stochastic attacks} that do not use controller internals, and (ii) \emph{policy-aware black-box attacks} that can query the deployed policy and observe actions or value estimates, but have no access to model parameters or gradients.

\noindent\textbf{Defense Assumption.} MDRC addresses mitigation after suspicious sensing has been flagged. We assume an \emph{orthogonal} anomaly-detection or sensor-health module triggers recovery, but we do not assume any oracle clean state, oracle attack label, or oracle knowledge of the perturbation type or magnitude. The detector only indicates that the current observation should not be trusted; MDRC is responsible for reconstructing a usable state for control. In our experiments, we apply MDRC directly to attacked observations to isolate recovery robustness from the accuracy of any particular detector.

\noindent\textbf{Attack Methods.} We instantiate the threat model with three classes of observation-space attacks that reflect both non-adaptive and adaptive adversaries. Firstly, we consider \emph{stochastic observation attacks}, which inject artificial noise into the observed state to mislead the TSC agent. This category includes:
\begin{itemize}
    \item \textbf{Gaussian Noise Attack:} The adversary injects Gaussian noise $\mathcal{N}(\mu, \sigma^2)$ scaled by intensity $k$ into state $\mathbf{s}_t$, yielding $\tilde{\mathbf{s}}_t = \mathbf{s}_t + k \cdot \mathcal{N}$. This models corruptions that can blend into natural sensor noise while steering the controller toward suboptimal states.
    \item \textbf{U-rand Attack:} The adversary introduces uniform random noise $\mathcal{U}$ within intensity $k$, expressed as $\tilde{\mathbf{s}}_t = \mathbf{s}_t + k \cdot \mathcal{U}(-\textbf{I}, \textbf{I})$, where $\textbf{I}$ is an all-one matrix. This models indiscriminate yet bounded perturbations to the observation channel.
\end{itemize}

Secondly, we consider \emph{adaptive policy-targeted attacks}, which use query access to craft perturbations that maximize control errors. These attacks include:
\begin{itemize}
    \item \textbf{MAD (Maximize Action Difference) Attack:} The adversary selects noise within an $\ell_\infty$ ball $\mathbf{B}_d(\mathbf{s}, k)$ to maximize the divergence between the current policy's output and its output on the perturbed state. Formally, $\tilde{\mathbf{s}}_t = \mathbf{s}_t + \arg\max_{\tilde{\mathbf{s}} \in \mathbf{B}_d(\mathbf{s}, k)} D(\pi_{\phi}(\cdot|\mathbf{s}) \parallel \pi_{\phi}(\cdot|\tilde{\mathbf{s}}))$. This adaptive attack pushes the controller toward action inconsistency.
    \item \textbf{MinQ (Minimize Q-value) Attack:} The adversary chooses perturbations within an $\ell_\infty$ ball $\mathbf{B}_d(\mathbf{s}, k)$ to minimize the $Q$-value for the chosen action or expected return, $\tilde{\mathbf{s}}_t = \mathbf{s}_t + \arg\min_{\tilde{\mathbf{s}} \in \mathbf{B}_d(\mathbf{s}, k)} Q(\tilde{\mathbf{s}}_t, \pi_{\phi}(\cdot|\tilde{\mathbf{s}}))$. This captures an adaptive attacker that exploits value estimates to induce low-quality control decisions.
\end{itemize}

Finally, we consider \emph{sensor availability attacks}, which model physical sensor damage or jamming that causes data loss. We represent this as $\tilde{\mathbf{s}}_t = \text{Mask} \cdot \mathbf{s}_t$, where masked dimensions correspond to unavailable sensor readings. This attack captures the loss of integrity and availability that forces the controller to operate under partial observability. Threat model captures realistic observation-space attacks in deployed cyber-physical control systems by considering black-box adversaries that manipulate the sensing pipeline without access to model internals.

\begin{figure*}[t]
    \centering
    \includegraphics[width=1\linewidth]{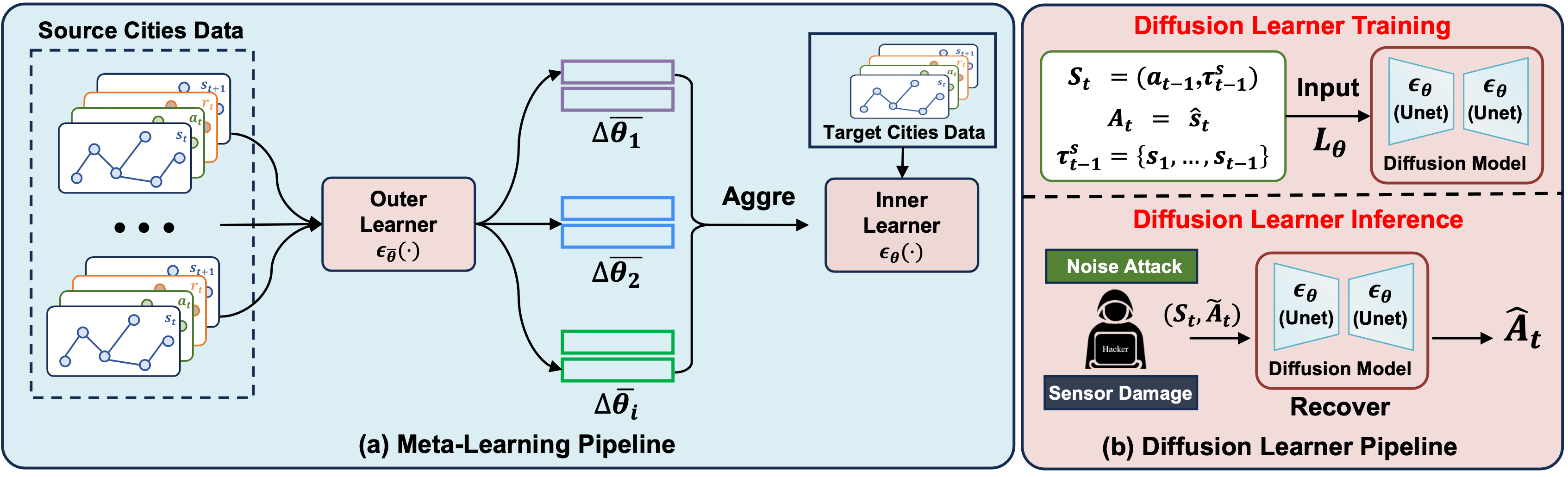}
    \caption{Overall framework of MDRC. MDRC is inserted as a plug-in recovery layer between sensing and the downstream controller. In the meta-learning pipeline, the outer diffusion learner acquires shared meta-parameters from source cities, while the inner learner adapts them to the target city. In the deployment pipeline, the trained recovery model reconstructs the state before control is executed.}
    \label{fig:framework}
\end{figure*}
\section{Methods}
MDRC is a post-detection state recovery defense mechanism. At runtime, an upstream detector or sensor-health monitor flags that the current observation may be corrupted; MDRC then reconstructs a trustworthy state vector $\hat{s}_t$ from the compromised observation $\tilde{s}_t$ and forwards $\hat{s}_t$ to the original controller. MDRC defends the sensing-to-control interface without modifying the downstream control logic. This design makes it compatible with both RL-based and rule-based TSC algorithms, as verified by our experiments. In addition, MDRC does not require an oracle clean state or attack label: its role is solely to mitigate corrupted sensing once recovery has been triggered.
\subsection{DDIM Recovery Dynamics and Reptile Meta-Learning}
We use DDIM and Reptile at two different levels of MDRC. Reptile operates in the parameter space of the recovery model and meta-learns an initialization that can be transferred across cities. DDIM operates in the state space at inference time and iteratively maps a corrupted observation toward a recovered traffic state. While these two procedures optimize different objects, the DDIM update admits a first-order optimization-like form that provides an intuitive connection between iterative state recovery and meta-learned adaptation. We use this connection as an interpretation of the design rather than as an equivalence between DDIM and Reptile. The
DDIM sampling step from time $t$ to $t-1$ can be written as:
\begin{align}
\mathbf{x}_{t-1} &= \frac{\sqrt{\bar{\alpha}_{t-1}}}{\sqrt{\bar{\alpha}_t}} \mathbf{x}_t \notag\\ 
&\quad - \left( \frac{\sqrt{\bar{\alpha}_{t-1}}\sqrt{1 - \bar{\alpha}_t}}{\sqrt{\bar{\alpha}_t}} - \sqrt{1 - \bar{\alpha}_{t-1} - \sigma_t^2} \right) \boldsymbol{\epsilon}_\theta^{(t)}(\mathbf{x}_t)+ \sigma_t \boldsymbol{\epsilon}\notag.
\label{eq:ddim-gda}
\end{align}

The following time-dependent parameters can be defined as: 
\begin{align*}
    \gamma &= \frac{\sqrt{\bar{\alpha}_{t-1}}}{\sqrt{\bar{\alpha}_t}}, \quad \xi = \sigma_t,\\
    \eta &= \frac{\sqrt{\bar{\alpha}_{t-1}}\sqrt{1 - \bar{\alpha}_t}}{\sqrt{\bar{\alpha}_t}}
           - \sqrt{1 - \bar{\alpha}_{t-1} - \sigma_t^2}.
\end{align*}

\noindent\textbf{Linking to Reptile Meta-Learning.} We begin by observing the structural similarity between the update rule in diffusion-based models and classical gradient-based optimization~\citep{zhang2024metadiff}. Consider the standard gradient descent formulation:
\begin{equation}
\theta \leftarrow \theta - \eta \nabla\mathcal{L}(\theta),
\label{eq_5}
\end{equation}
where $\theta$ denotes the model parameters, $\nabla\mathcal{L}(\theta)$ is the loss function, and $\eta$ is the learning rate. This iterative update aims to minimize the loss by moving $\theta$ in the direction of the negative gradient. In comparison, the deterministic update rule in DDIM can be reformulated as:
\begin{equation}
\mathbf{x}_{t-1} = \mathbf{x}_t - \eta\boldsymbol{\epsilon}_\theta^{(t)}(\mathbf{x}_t) + (\gamma - 1)\mathbf{x}_t + \xi\boldsymbol{\epsilon},
\label{eq_6}
\end{equation}
where $x_t$ denotes the generated sample at time step $t$, $\boldsymbol{\epsilon}_\theta^{(t)}$ is the learned noise predictor, and the additional terms $(\gamma - 1)\mathbf{x}_t$ and $\xi\boldsymbol{\epsilon}$ introduce momentum and stochasticity into the update dynamics, respectively. By comparing Eq.~(\ref{eq_5}) with Eq.~(\ref{eq_6}), it is observed that DDIM can be interpreted as performing a noise-conditioned, time-varying descent in the data space. $\boldsymbol{\epsilon}_\theta^{(t)}$ determines a gradient-like recovery direction, while $\eta$ remains a scaling factor modulating the update magnitude. This interpretation bridges the diffusion process with meta-learning. In particular, we draw parallels to the Reptile algorithm, a first-order meta-learning method that updates the meta-parameters $\theta$ by moving them toward task-specific adapted weights $\phi$, which are obtained by applying several gradient steps on a sampled task. Formally, Reptile performs: 
\begin{equation}
\theta \leftarrow \theta + \mu (\phi - \theta),
\label{eq:reptile}
\end{equation}
where $\phi$ approximates $\theta - \eta \nabla\mathcal{L}_\mathcal{T}(\theta)$ (with $\mathcal{T}$ being the specific task dataset) after a few steps gradient updates. The connection to Reptile is architectural rather than algebraic. During meta-training, Reptile optimizes the parameters $\theta$ of the diffusion recovery model across source-city tasks. For a sampled city $\mathcal{T}_i$, several inner updates produce  task-adapted parameters $\phi_i$, after which the shared initialization is updated toward $\phi_i$. DDIM uses the resulting meta-learned predictor $\epsilon_\theta$ to perform iterative recovery in the state space. Reptile improves the transferable initialization of the
recovery model, while DDIM provides the efficient inference-time recovery dynamics.

The additional term $(\gamma-1)\mathbf{x}_t$ introduces a time-dependent momentum-like effect, while $\xi\mathbf{\epsilon}$ controls sampling stochasticity. Both coefficients are determined by the diffusion schedule rather than learned as Reptile-style optimization parameters. These dynamics provide an efficient iterative recovery mechanism in the state space; cross-city adaptation is provided by Reptile through parameter-space meta-training.

DDIM extends beyond standard gradient descent by incorporating ideas from meta-optimization frameworks, such as Reptile. A key advantage is that important hyperparameters like $\gamma$ and $\xi$ are derived analytically from the diffusion schedule, rather than introduced as ad hoc heuristics. This gives a principled explanation for why the same diffusion process can support few-shot cross-city recovery while remaining efficient enough for deployment.

\subsection{Framework Overview}
We propose a meta-learning framework that unifies DDIM with Reptile updates for cross-city TSC. Each city is treated as a task in a multi-task paradigm: an \emph{outer loop} aggregates information across source cities and updates a shared initialization, while an \emph{inner loop} fine-tunes that initialization on a target city. At deployment time, the resulting model serves as a plug-in recovery module between sensing and control. Notably, both loops optimize the same diffusion-based loss, ensuring that the model learns a transferable state recovery mechanism rather than a city-specific denoiser.

\noindent\textbf{Meta-Learning Pipeline.}  Our diffusion learner is trained via offline meta-learning over a collection of multi-city datasets. We first collect logged trajectories $\mathcal{D}_\tau = \{(s_t, a_t, s_{t+1})\}$ from each training city $\tau$, and aggregate them into a global offline corpus $\mathcal{D}_{\text{meta}}$. During meta-training, the diffusion model is optimized to learn a cross-domain state distribution by minimizing the diffusion loss~\citep{yang2023dmbp}:
\begin{equation}
\begin{aligned}
  L_{\mathrm{diff}}(\theta; \mathcal{T}_i)
  = \mathbb{E}_{i \sim \mathcal{U}_K, \epsilon_t \sim \mathcal{N}(0,  I),(s_{t-N}, \ldots, s_{t+M-1}) \in D_\nu } \\
 \left\| \epsilon_{\theta}(\tilde{s}^i_t, c_{t-1},i) - \epsilon_{t}^i \right\|_{2}+
  \sum^{t+M-1}_{m=t+1}   \left\| \epsilon_{\theta}(\tilde{s}^i_m, \hat{c}_{m-1},i) - \epsilon_{m}^i \right\|_{2},  
\end{aligned}
\label{eq:diff-loss}
\end{equation}
where the condition represents as $c_t = (a_{t-1}, \tau^s_{t-1})$, $a_{t-1}$ is the previous TSC action, $\tau^s_{t-1}=\{s_1, ..., s_{t-1}\}$ is the TSC state trajectory, $\hat{c}_{m-1}=(a_{m-1},\tau^{\hat{s}}_{m-1})$, and $\tau^{\hat{s}}_{m-1}$ is the predicted state trajectory. This loss in a balances immediate and future timesteps by penalizing the mismatch between predicted and true noise across a window of length $N+M$.

\begin{algorithm}[t]
\caption{Cross-City Diffusion Meta-Learning}
\label{alg:diffusion-reptile}
\begin{algorithmic}[1]
\State \textbf{Input:} Source tasks $\{\mathcal{T}_i^{\mathrm{src}}\}$, target $\mathcal{T}_{\mathrm{tgt}}$, step sizes $\alpha,\eta$
\State Initialize $\theta$
\For{each outer iteration}
  \State $\phi \leftarrow 0$
  \For{each source task $\mathcal{T}_i^{\mathrm{src}}$}
    \State Compute $\phi \leftarrow \theta - \eta\,\nabla_\theta L_{\mathrm{diff}}(\theta;\mathcal{T}_i)$
    \State $\hat{\theta} \leftarrow \hat{\theta} + (\phi - \theta_i)$
  \EndFor
  \State $\theta \leftarrow \theta + \frac{\mu}{N}\,\sum_{i=1}^N\hat{\theta}$
\EndFor
\State Few-shot training $\theta$ on $\mathcal{T}_{\mathrm{tgt}}$ via $L_{\mathrm{diff}}$
\State \Return $\theta$
\end{algorithmic}
\end{algorithm}

We use Reptile to meta-learn an initialization $\theta$ that performs well on any city after a few updates. For each outer iteration with $N$ source tasks, we evaluate:
\begin{equation}
\begin{aligned}
     \phi
    = \theta - \eta \nabla_\theta L_{\mathrm{diff}}(\theta; \mathcal{T}_i),
    \quad 
    \theta \leftarrow \theta +  \frac{\mu}{N}\,\sum_{i=1}^N (\phi - \theta),
\end{aligned}
\end{equation}  
where $\mu$ is the outer-loop step size. This first-order update aligns the global parameters toward the average task-adapted parameters. During meta-training, we sample source tasks (e.g., $Hangzhou$) and run $k$ diffusion-loss inner updates to compute each $\phi$. The aggregated update $\theta$ refines the shared initialization. During meta-training, the denoising network is optimized on offline data from multiple source cities without environment interaction. After convergence, the diffusion model is frozen and deployed for state recovery in downstream control. At test time, the trained model supports zero-shot state reconstruction from noisy inputs. When transferring to a target city (e.g., $JiNan$), we further perform few-shot adaptation by fine-tuning $\theta$ with the diffusion loss (Eq.~\ref{eq:diff-loss}) to improve reconstruction fidelity and generalization. The algorithm is shown in Algorithm~\ref{alg:diffusion-reptile}. Our algorithm convergence demonstration is in theorem~\ref{thm:error_bound}. Then,

\begin{theorem}
\label{thm:error_bound}
Let $p_0$ denote the target clean-state distribution and $q_0$
denote the distribution induced by the discretized DDIM recovery
process. 
\begin{equation}
W_2(p_0,q_0)
\le
C_{\mathrm{score}}\sqrt{\mathcal{L}_{\mathrm{diff}}(\theta)}
+
C_{\mathrm{disc}}\Delta t
+
C_{\mathrm{init}}W_2(p_T,q_T),
\end{equation}
where $\Delta t$ is the discretization step of the reverse diffusion
dynamics. Under the regularity assumptions stated in Appendix~A.
\end{theorem}

\noindent\textbf{Remark.}
The bound characterizes the recovery process itself. It separates three sources of error: score-estimation error, numerical discretization error, and initialization mismatch. Reptile does not appear as a separate term in the bound; instead, its role is to meta-learn the parameters $\theta$ used by the score/noise predictor across source-city tasks. Therefore, better cross-city meta-training can reduce the first term indirectly by improving the transferred recovery model. Detailed proof is in Appendix~\ref{detialed_proof}.

\begin{algorithm}[t]
\caption{Repaint algorithm of MDRC}
\label{alg:repaint}
\begin{algorithmic}[1]
    \State Input $\tilde{s}_t^{j}, c_{t-1},m$
\For{$j=1$ to $K$}
  \For{$u=1$ to $U$}
     \State $\epsilon \sim \mathcal{N}(0,I)$ if $j>1$, else $\epsilon=0$
     \State Get $\tilde{s}_{t,known}^{j-1}$ by Equation (10)
     \State $z \sim \mathcal{N}(0,I)$ if $j>1$, else $z=0$
     \State Get $\tilde{s}^{j-1}_{t,unknown}$ by Equation (11)
     \State Get recovered $\tilde{s}^{j-1}_t$ by Equation (12)
     \If {$u<U$ and $j>1$}
     \State   $\tilde{s}^j_t \sim \mathcal{N}(\sqrt{1-\beta_{j-1}}\tilde{s}^{j-1}_t,\beta_{j-1}I)$
     \EndIf
    \EndFor
 \EndFor
\State Return $\hat{s}_t$
 \end{algorithmic}
\end{algorithm}

\noindent\textbf{Diffusion Pipeline.} The pipeline starts from \textit{Denoising via DDIM.} During evaluation, TSC sensors may be corrupted by Gaussian noise, MAD, U-rand, or Min-Q perturbations. To restore the true state, we employ the diffusion model meta-trained by Reptile. At each diffusion timestep \(j\), the network receives the current noisy observation \(\tilde{s}_t^j\), the last estimated state \(c_{t-1}\), and the timestep index \(j\). 
The denoising update follows the DDIM rule:
\begin{equation}
\begin{split} 
\tilde{s}_t^{\,j-1}
&= \sqrt{\bar\alpha_{\,j-1}}\,
    \frac{\tilde{s}_t^j - \sqrt{1-\bar\alpha_j}\,\epsilon_{\theta}(\tilde{s}_t^j,\,c_{t-1},\,j)}
         {\sqrt{\bar\alpha_j}} \\ 
  &\quad + \sqrt{1-\bar\alpha_{\,j-1}-\sigma_j^2}\;\epsilon_{\theta}(\tilde{s}_t^j,\,c_{t-1},\,j) \\ 
  &\quad + \sigma_j z\,,
\end{split}
\end{equation}
where \(z\!\sim\!\mathcal{N}(0,I)\). Iterating this process for \(j= T,..., 1\) yields the reconstructed state \(\hat{s}_t\), effectively removing adversarial or stochastic corruptions before the controller consumes the state.

Next, \textit{Repainting} is conducted to compensate the missing parts. We adapt conditional diffusion to infer and restore missing or damaged sensor readings. Let \(m\) be the binary mask indicating known (\(m=1\)) and unknown (\(m=0\)) entries in \(\tilde{A}_t^{\,j}\). At each reverse step, we sample the known portion by forward diffusion:
\begin{equation}
    \tilde{s}_{t,\text{known}}^{\,j-1}
    = \sqrt{\bar\alpha_j}\,\tilde{s}_{t,\text{known}}
      + \sqrt{1-\bar\alpha_j}\,\epsilon,
\end{equation}
and recover the unknown portion via the conditional reverse update:
\begin{equation}
\begin{split} 
\tilde{s}_{t,\,\text{unknown}}^{\,j-1}
&= \sqrt{\bar\alpha_{\,j-1}}\,
    \frac{\tilde{s}_{t,\,\text{unknown}}^j
          - \sqrt{1-\bar\alpha_j}\,\epsilon_{\theta}(\tilde{s}_t^j,\,c_{t-1},\,j)}
         {\sqrt{\bar\alpha_j}} \\ 
  &\quad + \sqrt{1 - \bar\alpha_{\,j-1} - \sigma_j^2}\;\epsilon_{\theta}(\tilde{s}_t^j,\,c_{t-1},\,j)
  + \sigma_j\,z,
\end{split}
\end{equation}
where \(\sigma\) and \(z\) are independent Gaussian noises. The combined sample for the next iteration is
\begin{equation}
    \tilde{s}_t^{\,j-1}
    = m \odot \tilde{s}_{t,\text{known}}^{\,j-1}
      + (1-m)\odot \tilde{s}_{t,\text{unknown}}^{\,j-1}.
\end{equation}
By iterating from \(j=T\) down to \(j=1\), the repaint algorithm reconstructs the full state \(\hat{s}_t\). In both denoising and repainting, MDRC acts as the same post-detection recovery layer and leaves the downstream controller unchanged. The complete procedure is detailed in Algorithm~\ref{alg:repaint}.

\subsection{Detailed Algorithms}

The Cross-City Diffusion Meta-Learning algorithm, outlined in Algorithm~\ref{alg:diffusion-reptile}, is designed to enable effective knowledge transfer across multiple city-specific tasks for diffusion-based models. By leveraging a meta-learning framework inspired by Reptile, the algorithm initializes a shared parameter set $\theta$ and iteratively updates it using gradients computed from source tasks $\mathcal{T}_i^{\mathrm{src}}$. For each source task, it performs an inner update with step size $\eta$ to compute task-specific parameters $\theta_i'$, aggregates the parameter differences, and updates the global parameters using a meta-learning rate $\mu$. After training on source tasks, the model is fine-tuned on the target task $\mathcal{T}_{\mathrm{tgt}}$ to adapt to specific city characteristics, enhancing generalization in diffusion-based applications such as urban data modeling.

The Repaint Algorithm of MDRC, presented in Algorithm~\ref{alg:repaint}, is a robust iterative method for reconstructing signals in diffusion-based models, particularly suited for tasks requiring inpainting or signal recovery under noisy conditions. The algorithm takes as input an initial signal estimate $\tilde{s}_t^{i}$, a context $c_{t-1}$, and a mask $m$, and performs $K$ iterations, each with $U$ inner steps. In each step, it samples noise $\epsilon$ and $z$ from a standard normal distribution (except in the first iteration, where noise is set to zero) and applies Equations (10), (11), and (12) to update known and unknown signal components and recover the signal. A stochastic update is applied when necessary, governed by the parameter $\beta_{i-1}$, to introduce controlled noise, ultimately producing a refined signal estimate $\hat{s}_t$ that enhances robustness in applications like image or data reconstruction.

\makeatletter
\@ifundefined{color@theorycol}{\definecolor{theorycol}{RGB}{220,232,243}}{}
\makeatother
\newlength{\evalblockwidth}
\setlength{\evalblockwidth}{0.96\columnwidth}
\newenvironment{evalblock}{%
    \par\smallskip
    \begin{center}
    \begin{minipage}{\evalblockwidth}
    \noindent
}{%
    \end{minipage}
    \end{center}
    \par\smallskip
}
\newcommand{\takeawaybox}[2]{%
    \par\smallskip
    \begin{center}
    {%
        \setlength{\fboxsep}{6pt}%
        \colorbox{theorycol!25}{%
            \begin{minipage}{\dimexpr\evalblockwidth-2\fboxsep\relax}
            \textbf{Takeaway (#1).} #2
            \end{minipage}
        }%
    }%
    \end{center}
    \par\smallskip
}
\section{REAL-WORLD SENSOR-FAILURE AND HARDWARE-IN-THE-LOOP VALIDATION}
\label{sec:real_world_deployment}
\begin{figure}[h]
    \centering

    \begin{minipage}[c]{0.58\linewidth}
        \centering
        \includegraphics[width=\linewidth]{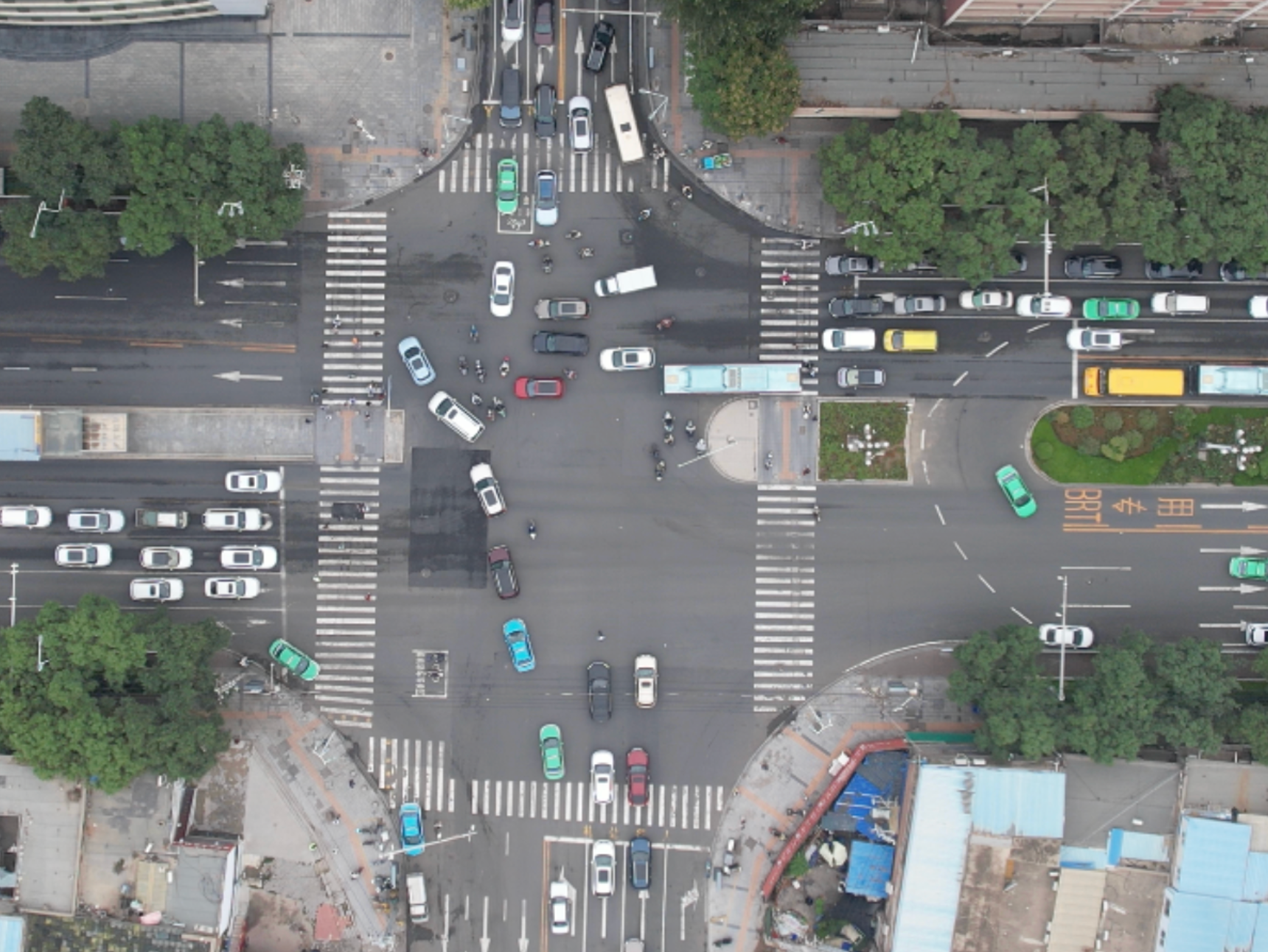}
        \vspace{-1mm}
        {\small\textbf{(a)}}
    \end{minipage}
    \hfill
    \begin{minipage}[c]{0.33\linewidth}
        \centering
        \includegraphics[width=\linewidth]{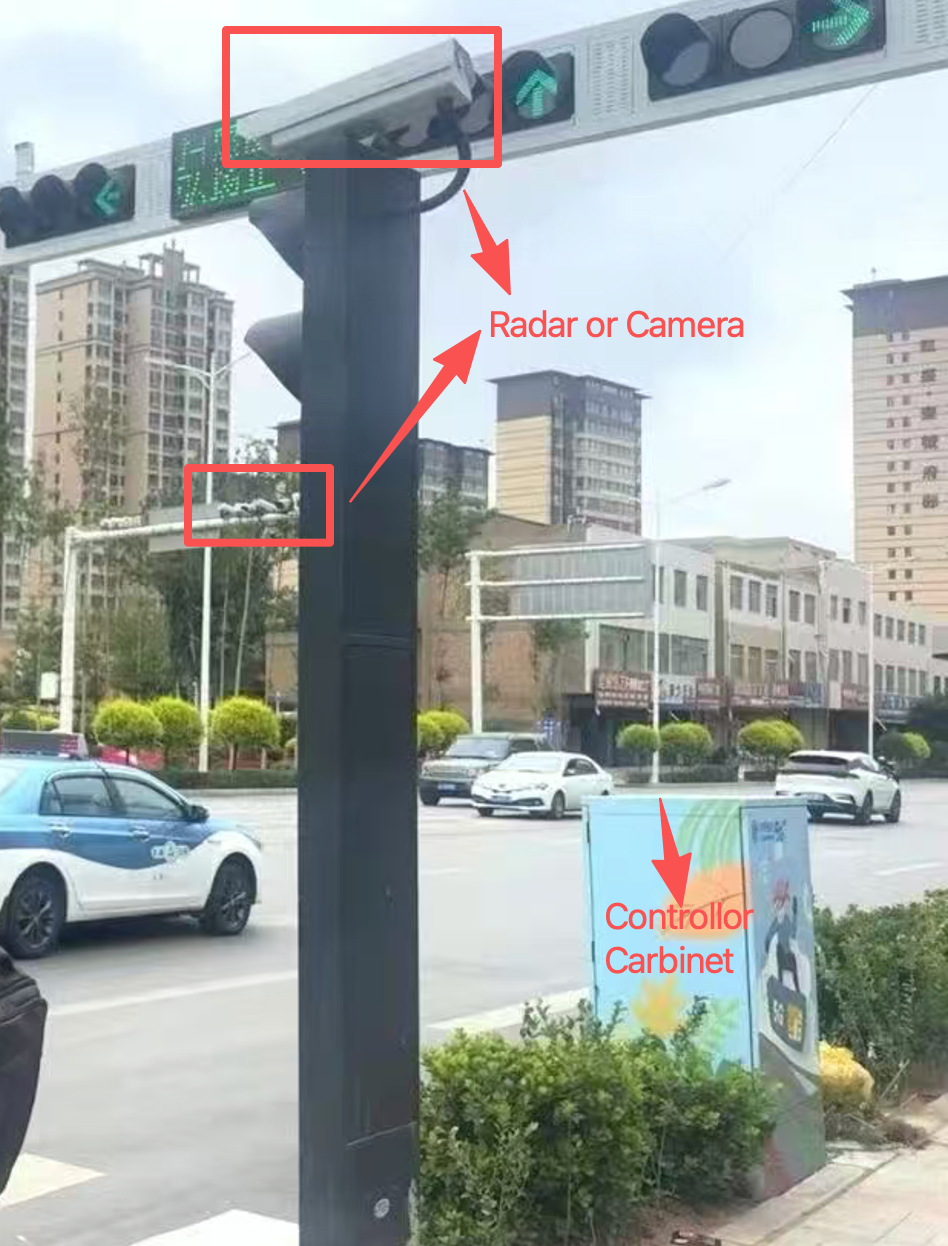}
        \vspace{-1mm}
        {\small\textbf{(b)}}
    \end{minipage}

    \vspace{-1mm}
    \caption{\textbf{Real-world deployment environment and roadside infrastructure.}
    (a) Operational urban intersection used for real-world sensing and system validation.
    (b) Physical roadside sensing and signal-control infrastructure at the deployment site.}
    \label{fig:realworld_deployment}
\end{figure}
Figure~\ref{fig:realworld_deployment} shows the operational
intersection and the physical roadside sensing and signal-control
infrastructure used in our real-world validation.

We conduct two complementary experiments to evaluate whether MDRC can
operate in a real-world TSC system. First, we conduct a controlled sensor-failure experiment using measurements collected from an operational roadside detector system. We emulate a severe availability failure by disabling 50\% of the detector channels and evaluate whether MDRC can reconstruct the unavailable sensing information from the remaining physical measurements. This experiment preserves the temporal dynamics and cross-channel correlations of genuine roadside sensing while providing a controlled ground truth for quantifying recovery quality. Second, we integrate MDRC into a hardware-in-the-loop
(HIL) traffic-signal stack containing physical roadside sensors, an edge computer, a central inference server, and a commercial signal controller. These experiments evaluate real-data recovery and prototype-level deployment feasibility, respectively.

\paragraph{Controlled sensor-failure recovery on real roadside measurements}
We use a morning-peak record collected at an anonymized city's road intersection. The trace contains 3,600 consecutive one-second observations from twelve movement-level detector channels. To emulate severe sensor unavailability, controlled 50\% channel unavailability/dropout.
\begin{figure}[h]
    \centering
    \includegraphics[width=\columnwidth]
    {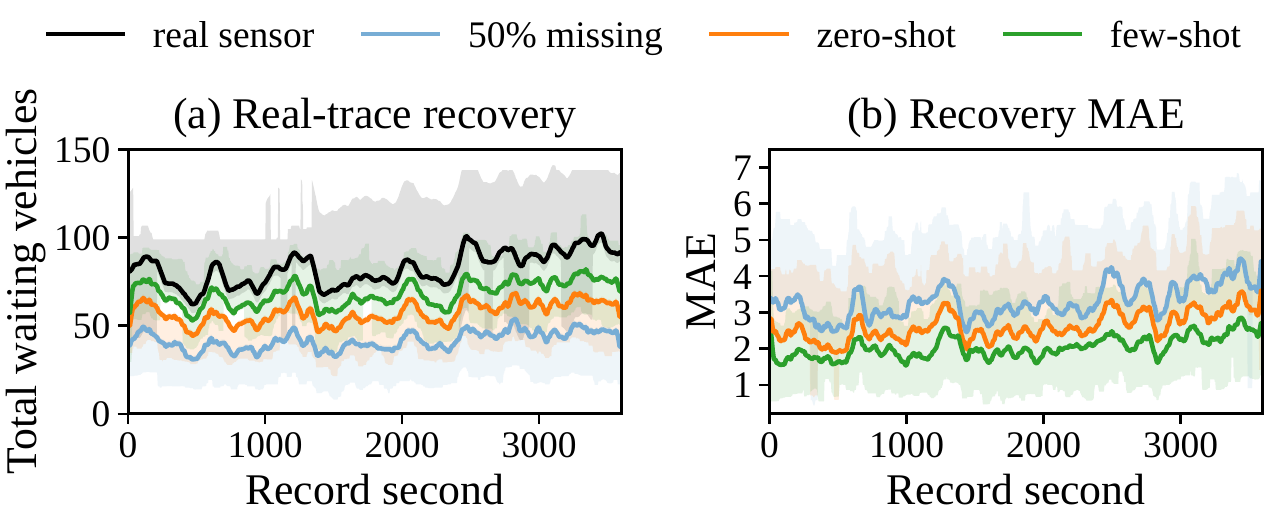}
    \caption{Recovery on a measured roadside-detector trace with
    controlled 50\% sensor-availability failure on real roadside measurements.}
    \label{fig:qihu_real_data_recovery}
\end{figure}
MDRC reconstructs the unavailable channels using the remaining
measurements, the preceding four causal states, and the previous signal
phase; no future observations are used. As shown in
Fig.~\ref{fig:qihu_real_data_recovery}, both zero-shot and few-shot
MDRC recover substantially more of the measured temporal structure
than the unrecovered 50\%-missing input, with few-shot adaptation further reducing reconstruction error.

\paragraph{Hardware-in-the-loop integration}
The HIL testbed is deployed at an anonymized city's road intersection. Four directional roadside radars provide timestamped
vehicle detections, which are converted into lane-level running and
waiting vehicle states by an RK3588 industrial edge computer. The
states and sensor-health information are transmitted through
fiber-optic Ethernet to a server equipped with a single RTX~4090.

MDRC is inserted at the state interface before the deployed downstream
TSC algorithm service runs FuzzyLight~\cite{li2025fuzzylightrobusttwostagefuzzy}. MDRC is inserted before FuzzyLight at the lane-state interface and
does not modify its policy parameters, action space, or signal plan.
The commercial signal controller remains responsible for enforcing
the operational phase and timing constraints. It reconstructs unreliable channels and forwards the
recovered state to the unchanged controller. The resulting phase
recommendation is returned to the edge device, which validates the
control status, phase identifier, and signal-timing constraints before
forwarding it to the commercial signal controller. Videos of the
physical sensing and controller-in-the-loop stack are included in the
anonymous artifact.

\paragraph{Runtime and control validity}
The HIL runtime log validates the continuously deployed
sensing-to-controller path, while MDRC inference is benchmarked
separately on the same class of RTX~4090 server. Combining the two
measured components gives an estimated processing latency of
approximately 38~ms, which occupies 3.8\% of the one-second control
interval.

MDRC inference takes approximately 28~ms on a single RTX~4090. Together
with the approximately 10~ms existing control path, the component-wise
latency is approximately 38~ms, occupying 3.8\% of the one-second
control interval.

\begin{table}[t]
    \centering
    \caption{Hardware-in-the-loop runtime and control-validity results.}
    \label{tab:hil_runtime}
    \begin{tabular}{lr}
        \hline
        Metric & Result \\
        \hline
        Runtime duration & 9.16~h \\
        Sensing/control cycles & 32,389 \\
        Decision availability & 99.79\% \\
        Valid recommendations & 32,319 / 32,388 \\
        Out-of-plan recommendations & 0 \\
        MDRC inference latency & $\sim$28~ms \\
        Combined component latency & $\sim$38~ms \\
        Fraction of one-second budget & 3.8\% \\
        \hline
    \end{tabular}
\end{table}

\takeawaybox{HIL}{
Under a controlled 50\% sensor-availability failure, MDRC reconstructs unavailable channels directly from genuine roadside measurements without simulator-generated observations. The complementary HIL experiment further verifies that the same recovery component can operate within a physical sensing-to-control stack comprising roadside radars, edge acquisition, fiber communication, GPU inference, and a commercial signal-controller interface. }

\section{Experiments}

All experiments were conducted on an Ubuntu 22.04 server equipped with 8 x NVIDIA GeForce RTX 4090 GPUs and 512GB DDR5 RAM. However, both the training and inference of MDRC only require a single RTX 4090 GPU.

\subsection{Datasets and Compared Methods}
\label{subsec:datasets_methods}

We evaluate our approach using real-world traffic datasets simulated in Cityflow~\citep{CityFLow}, measuring the Average Travel Time (ATT) over a 60-minute period. The datasets include start/end vehicle points with a fixed motion model, covering 7 traffic datasets from JiNan and HangZhou (China) and New York~\citep{FRAP2019} (USA). JiNan includes 12 intersections ($3\times4$ grid) with three datasets: $JiNan_1$, $JiNan_2$, and $JiNan_3$. HangZhou has 16 intersections ($4\times4$ grid) and two datasets: $HangZhou_1$ and $HangZhou_2$. New York features a large-scale network with 192 intersections ($28\times7$ grid) and two datasets: $Newyork_1$ and $Newyork_2$.

We compare our method against traditional and RL-based TSC approaches. Traditional methods include FixedTime~\citep{webster1958} and Advanced-Maxpressure~\citep{icml}. RL-based methods include Advanced-CoLight~\citep{icml} and Advanced-Mplight~\citep{icml}, which is based on FRAP~\citep{FRAP2019}, and RobustLight~\citep{lirobustlight}. MDRC is attached as a recovery layer to these controllers without changing their internal control logic, which allows us to evaluate controller-agnostic deployment. Our evaluation covers stochastic attacks (Gaussian, U-rand), adaptive policy-targeted attacks (MAD, MinQ), and sensor failures, typically at two perturbation strengths to assess robustness under increasing attack power. All results are averaged over ten independent runs unless otherwise stated. The complete hyperparameter configuration used for model training is provided in Appendix~\ref{hyperparameter}.

\subsection{Research Questions}
We organize the experimental section around six research questions that reflect a security evaluation mindset for resilient CPS defense:
\begin{itemize}
    \item \textbf{RQ1.} Can MDRC defend against both stochastic and adaptive observation attacks, especially as attack strength increases?
    \item \textbf{RQ2.} Can MDRC maintain control quality under sensor failures and structured missingness?
    \item \textbf{RQ3.} Does MDRC transfer to unseen cities under attack in zero-shot and few-shot settings?
    \item \textbf{RQ4.} Which components of MDRC are necessary for the robustness-latency trade-off?
    \item \textbf{RQ5.} Is MDRC practical as an online plug-in defense in terms of latency and resource cost?
    \item \textbf{RQ6.} Does MDRC improve state recovery fidelity beyond ATT under broader missingness conditions?
\end{itemize}

\begin{table*}[t]
\caption{Performance of ATT in $JiNan$, $HangZhou$. Our MDRC recovers the state of traditional and RL-based TSC algorithms.}
\label{table:noiseattack}
\centering
   \fontsize{7}{8}\selectfont
\renewcommand{\arraystretch}{1.5}
\setlength{\tabcolsep}{4pt}

\begin{tabular}{cccccccccc} 
\toprule
\multirow{2}{*}{Dataset}      & \multirow{2}{*}{\begin{tabular}[c]{@{}c@{}}Noise \\Type\end{tabular}} & \multirow{2}{*}{\begin{tabular}[c]{@{}c@{}}~Noise \\Scale\end{tabular}} & FixedTime               & \multicolumn{3}{c}{Advanced-CoLight}                                                                                                                                 & \multicolumn{3}{c}{Advanced-MpLight}                                                                                                          \\ 
\cline{4-10}
                              &                                                                       &                                                                         & base                    & base                            & RobustLight                                                       & MDRC                                                      & base                  & RobustLight                                              & MDRC                                              \\ 
\hline
\multirow{8}{*}{$JiNan_1$}    & \multirow{2}{*}{Gaussian}                                             & 3.5                                                                     & \multirow{8}{*}{428.11} & 316.96±4.63   & 328.38±3.90                                     & {\cellcolor[rgb]{0.753,0.753,0.753}}\textbf{\textbf{282.26±1.35}}  & 327.93\textbf{±}8.76  & 303.83\textbf{±}5.98                                     & {\cellcolor[rgb]{0.753,0.753,0.753}}\textbf{289.85±3.76}   \\
                              &                                                                       & 4.0                                                                     &                         & 329.24\textbf{\textbf{±}}3.73   & 346.47\textbf{\textbf{±}}3.85                                     & {\cellcolor[rgb]{0.753,0.753,0.753}}\textbf{\textbf{281.73±1.77}}  & 338.68\textbf{±}7.41  & 307.19\textbf{±}2.21                                     & {\cellcolor[rgb]{0.753,0.753,0.753}}\textbf{289.25±1.82}   \\
                              & \multirow{2}{*}{U-rand}                                               & 3.5                                                                     &                         & 483.26\textbf{\textbf{±}}74.90  & 365.55\textbf{\textbf{±}}11.95                                    & {\cellcolor[rgb]{0.753,0.753,0.753}}\textbf{\textbf{327.33±3.81}}  & 417.81\textbf{±}12.34 & {\cellcolor[rgb]{0.753,0.753,0.753}}\textbf{362.61±9.16} & 399.88\textbf{±}19.17                                      \\
                              &                                                                       & 4.0                                                                     &                         & 497.32\textbf{\textbf{±}}64.04  & 388.69\textbf{\textbf{±}}25.74                                    & {\cellcolor[rgb]{0.753,0.753,0.753}}\textbf{\textbf{329.89±8.66}}  & 434.84\textbf{±}11.18 & {\cellcolor[rgb]{0.753,0.753,0.753}}\textbf{365.96±7.52} & 377.99\textbf{±}16.26                                      \\
                              & \multirow{2}{*}{MAD}                                                  & 3.5                                                                     &                         & 454.77\textbf{\textbf{±}}9.38   & 301.29\textbf{\textbf{±}}3.43                                     & {\cellcolor[rgb]{0.753,0.753,0.753}}\textbf{\textbf{282.45±2.03}}  & 339.41\textbf{±}37.82 & 305.16\textbf{±}19.99                                    & {\cellcolor[rgb]{0.753,0.753,0.753}}\textbf{279.95±1.69}   \\
                              &                                                                       & 4.0                                                                     &                         & 495.47\textbf{\textbf{±}}15.55  & 305.32\textbf{\textbf{±}}4.13                                     & {\cellcolor[rgb]{0.753,0.753,0.753}}\textbf{\textbf{286.63±2.31}}  & 355.38\textbf{±}36.38 & 313.24\textbf{±}30.45                                    & {\cellcolor[rgb]{0.753,0.753,0.753}}\textbf{290.00±1.61}   \\
                              & \multirow{2}{*}{MinQ}                                                 & 3.5                                                                     &                         & 493.10\textbf{\textbf{±}}96.80  & 388.41\textbf{\textbf{±}}85.49                                    & {\cellcolor[rgb]{0.753,0.753,0.753}}\textbf{\textbf{299.67±8.34}}  & 347.62\textbf{±}28.10 & 297.97\textbf{±}14.57                                    & {\cellcolor[rgb]{0.753,0.753,0.753}}\textbf{284.94±7.45}   \\
                              &                                                                       & 4.0                                                                     &                         & 520.99\textbf{\textbf{±}}115.29 & 387.87\textbf{\textbf{±}}62.70                                    & {\cellcolor[rgb]{0.753,0.753,0.753}}\textbf{\textbf{286.29±2.57}}  & 356.30\textbf{±}30.91 & 307.07\textbf{±}21.89                                    & {\cellcolor[rgb]{0.753,0.753,0.753}}\textbf{288.30±4.54}   \\ 
\hline
\multirow{8}{*}{$JiNan_3$}    & \multirow{2}{*}{Gaussian}                                             & 3.5                                                                     & \multirow{8}{*}{383.01} & 320.04\textbf{\textbf{±}}11.40   & 277.17\textbf{\textbf{±}}2.40                                     & {\cellcolor[rgb]{0.753,0.753,0.753}}\textbf{\textbf{263.89±1.92}}  & 378.13\textbf{±}32.17 & 279.79\textbf{±}1.79                                     & {\cellcolor[rgb]{0.753,0.753,0.753}}\textbf{261.91±1.17}   \\
                              &                                                                       & 4.0                                                                     &                         & 331.64\textbf{\textbf{±}}14.13  & 283.54\textbf{\textbf{±}}4.32                                     & {\cellcolor[rgb]{0.753,0.753,0.753}}\textbf{\textbf{259.75±2.31}}  & 390.93\textbf{±}25.46 & 290.37\textbf{±}2.56                                     & {\cellcolor[rgb]{0.753,0.753,0.753}}\textbf{270.69±1.45}   \\
                              & \multirow{2}{*}{U-rand}                                               & 3.5                                                                     &                         & 481.55\textbf{\textbf{±}}43.52  & 333.46\textbf{\textbf{±}}19.05                                    & {\cellcolor[rgb]{0.753,0.753,0.753}}\textbf{\textbf{290.39±1.81}}  & 496.63\textbf{±}44.43 & 354.45\textbf{±}11.06                                    & {\cellcolor[rgb]{0.753,0.753,0.753}}\textbf{314.14±5.68}   \\
                              &                                                                       & 4.0                                                                     &                         & 490.09\textbf{\textbf{±}}44.06  & 344.34\textbf{\textbf{±}}14.80                                    & {\cellcolor[rgb]{0.753,0.753,0.753}}\textbf{\textbf{288.89±3.16}}  & 505.93\textbf{±}41.98 & 363.10\textbf{±}18.75                                    & {\cellcolor[rgb]{0.753,0.753,0.753}}\textbf{312.17±3.62}   \\
                              & \multirow{2}{*}{MAD}                                                  & 3.5                                                                     &                         & 446.89\textbf{\textbf{±}}15.76  & {\cellcolor[rgb]{0.753,0.753,0.753}}\textbf{\textbf{273.78±4.00}} & 281.70\textbf{\textbf{±}}3.29                                      & 442.19\textbf{±}81.19 & 268.96\textbf{±}11.41                                    & {\cellcolor[rgb]{0.753,0.753,0.753}}\textbf{259.25±2.87}   \\
                              &                                                                       & 4.0                                                                     &                         & 478.85\textbf{\textbf{±}}22.27  & {\cellcolor[rgb]{0.753,0.753,0.753}}\textbf{\textbf{276.64±2.32}} & 298.49\textbf{\textbf{±}}3.37                                      & 446.93\textbf{±}27.17 & 279.83\textbf{±}18.33                                    & {\cellcolor[rgb]{0.753,0.753,0.753}}\textbf{279.05±2.91}   \\
                              & \multirow{2}{*}{MinQ}                                                 & 3.5                                                                     &                         & 376.00\textbf{\textbf{±}}24.65  & 298.31\textbf{\textbf{±}}20.34                                    & {\cellcolor[rgb]{0.753,0.753,0.753}}\textbf{\textbf{283.16±4.16}}  & 423.76\textbf{±}50.93 & 278.75\textbf{±}17.00                                    & {\cellcolor[rgb]{0.753,0.753,0.753}}\textbf{277.32±17.16}  \\
                              &                                                                       & 4.0                                                                     &                         & 406.03\textbf{\textbf{±}}6.45   & 315.39\textbf{\textbf{±}}18.80                                    & {\cellcolor[rgb]{0.753,0.753,0.753}}\textbf{\textbf{286.90±2.48}}  & 465.95\textbf{±}46.43 & 297.21\textbf{±}23.12                                    & {\cellcolor[rgb]{0.753,0.753,0.753}}\textbf{288.77±8.31}   \\ 
\hline
\multirow{8}{*}{$HangZhou_2$} & \multirow{2}{*}{Gaussian}                                             & 3.5                                                                     & \multirow{8}{*}{406.65} & 495.92±23.47 & 353.49±6.46 & {\cellcolor[rgb]{0.753,0.753,0.753}}\textbf{343.31±4.65} & 429.53±13.96 & 367.62±8.48 & {\cellcolor[rgb]{0.753,0.753,0.753}}\textbf{352.33±8.51}
   \\
                              &                                                                       & 4.0                                                                     &                         & 520.59±17.34 & 355.93±5.86 & {\cellcolor[rgb]{0.753,0.753,0.753}}\textbf{339.43±3.60} & 432.60±6.00 & 378.09±9.33 & {\cellcolor[rgb]{0.753,0.753,0.753}}\textbf{359.60±11.89}
   \\
                              & \multirow{2}{*}{U-rand}                                               & 3.5                                                                     &                         & 567.56±20.09 & 415.49±11.33 & {\cellcolor[rgb]{0.753,0.753,0.753}}\textbf{364.17±4.22} & 481.32±37.20 & 426.04±13.45 & {\cellcolor[rgb]{0.753,0.753,0.753}}\textbf{402.33±10.19}
                                       \\
                              &                                                                       & 4.0                                                                     &                         & 566.64±17.55 & 429.26±11.82 & {\cellcolor[rgb]{0.753,0.753,0.753}}\textbf{370.33±3.10} & 472.05±42.80 & 433.15±9.94 & {\cellcolor[rgb]{0.753,0.753,0.753}}\textbf{397.78±9.88}
                                       \\
                              & \multirow{2}{*}{MAD}                                                  & 3.5                                                                     &                         & 496.73±22.83 & {\cellcolor[rgb]{0.753,0.753,0.753}}\textbf{333.93±3.71} & 341.95±6.63 & 433.46±32.89 & 362.01±6.67 & {\cellcolor[rgb]{0.753,0.753,0.753}}\textbf{350.77±9.28}
   \\
                              &                                                                       & 4.0                                                                     &                         & 528.74±28.18 & {\cellcolor[rgb]{0.753,0.753,0.753}}\textbf{339.41±8.00} & 344.82±2.94 & 471.26±29.52 & {\cellcolor[rgb]{0.753,0.753,0.753}}\textbf{363.45±14.04} & 370.91±15.03
                                       \\
                              & \multirow{2}{*}{MinQ}                                                 & 3.5                                                                     &                         & 441.72±31.86 & {\cellcolor[rgb]{0.753,0.753,0.753}}\textbf{345.80±2.82} & 346.51±5.91 & 425.09±30.34 & {\cellcolor[rgb]{0.753,0.753,0.753}}\textbf{356.49±5.33} & 361.75±6.50
                                       \\
                              &                                                                       & 4.0                                                                     &                         & 478.9±20.02 & {\cellcolor[rgb]{0.753,0.753,0.753}}\textbf{349.58±7.40} & 350.15±4.85 & 450.80±31.37 & {\cellcolor[rgb]{0.753,0.753,0.753}}\textbf{363.12±6.40} & 369.61±5.97
                                      \\
\bottomrule
\end{tabular}

\end{table*}

\subsection{Results}
\label{subsec:overall_results}

We answer these research questions below. Rather than reporting average performance only, we emphasize robustness under stronger attacks, behavior under adaptive attackers, cross-city transfer under attack, and deployment cost.

\noindent\textbf{RQ1: Can MDRC defend against both stochastic and adaptive observation attacks?}
\label{subsubsec:noise_attack_state}

Table~\ref{table:noiseattack} evaluates MDRC under both stochastic attacks (Gaussian, U-rand) and adaptive policy-targeted attacks (MAD, MinQ), each at two perturbation strengths. This provides an attack-strength sweep and lets us test whether the defense remains effective beyond a single corruption pattern. MDRC improves average performance and provides substantial gains in many high-corruption regimes. On average, MDRC achieves 6.77\% lower ATT, with the largest improvement reaching 26.18\% under the adaptive MinQ attack on $JiNan_1$ with Advanced-CoLight. These results support the claim that MDRC acts as a state recovery defense rather than a noise-specific heuristic.

\begin{table}[h]
\caption{Performance of ATT in $JiNan$, $HangZhou$. Our MDRC recovers the state of traditional and RL-based TSC algorithms to evaluate the performance.}
\label{table:noiseattack1}
\centering
   \fontsize{7}{8}\selectfont
\renewcommand{\arraystretch}{1.5}
\setlength{\tabcolsep}{4pt}
\begin{tabular}{cccccc} 
\toprule
\multirow{2}{*}{Dataset}      & \multirow{2}{*}{\begin{tabular}[c]{@{}c@{}}Noise \\Type\end{tabular}}  
& \multirow{2}{*}{\begin{tabular}[c]{@{}c@{}}~Noise \\Scale\end{tabular}}  & \multicolumn{3}{c}{Advanced-MaxPressure}                                                                                                                               \\ 
\cline{4-6}
                              &             &               & base                          & RobustLight                                                       & MDRC                                                      \\ 
\hline
\multirow{4}{*}{$JiNan_1$}    & \multirow{2}{*}{Gaussian}    & 3.5
                              & 285.72\textbf{\textbf{±}}2.20 & 283.47\textbf{\textbf{±}}1.16                                     & {\cellcolor[rgb]{0.753,0.753,0.753}}\textbf{\textbf{279.85±1.20}}  \\
                              &                              & 4.0
                              & 289.74\textbf{\textbf{±}}2.61 & 288.90\textbf{\textbf{±}}1.18                                     & {\cellcolor[rgb]{0.753,0.753,0.753}}\textbf{\textbf{283.95±1.30}}  \\
                              & \multirow{2}{*}{U-rand}      & 3.5
                              & 312.67\textbf{\textbf{±}}2.81 & {\cellcolor[rgb]{0.753,0.753,0.753}}\textbf{\textbf{307.08±2.67}} & 307.55\textbf{\textbf{±}}2.07                                      \\
                              &                              & 4.0
                              & 321.26\textbf{\textbf{±}}1.81 & {\cellcolor[rgb]{0.753,0.753,0.753}}\textbf{\textbf{312.52±2.82}} & 334.73\textbf{\textbf{±}}1.70                                      \\ 
\hline
\multirow{4}{*}{$JiNan_3$}    & \multirow{2}{*}{Gaussian}    & 3.5
                               & 268.67\textbf{\textbf{±}}1.56 & 268.69\textbf{\textbf{±}}1.32                                    & {\cellcolor[rgb]{0.753,0.753,0.753}}\textbf{\textbf{262.13±0.60}}  \\
                               &                              & 4.0
                              & 273.22\textbf{\textbf{±}}0.77 & 272.95\textbf{\textbf{±}}1.68                                     & {\cellcolor[rgb]{0.753,0.753,0.753}}\textbf{\textbf{265.42±1.56}}  \\
                              & \multirow{2}{*}{U-rand}      & 3.5
                              & 293.44\textbf{\textbf{±}}2.74 & 291.58\textbf{\textbf{±}}1.74                                     & {\cellcolor[rgb]{0.753,0.753,0.753}}\textbf{\textbf{285.97±1.11}}  \\
                              &                              & 4.0
                              & 300.53\textbf{\textbf{±}}3.67 & 296.06\textbf{\textbf{±}}2.23                                     & {\cellcolor[rgb]{0.753,0.753,0.753}}\textbf{\textbf{295.45±1.85}}  \\ 
\hline
\multirow{4}{*}{$HangZhou_2$} & \multirow{2}{*}{Gaussian}    & 3.5
                              & 345.87\textbf{\textbf{±}}1.37 & 346.53\textbf{\textbf{±}}1.61                                     & {\cellcolor[rgb]{0.753,0.753,0.753}}\textbf{\textbf{341.09±1.08}}  \\
                              &                              & 4.0
                              & 348.39\textbf{\textbf{±}}2.53 & 350.68\textbf{\textbf{±}}1.44                                     & {\cellcolor[rgb]{0.753,0.753,0.753}}\textbf{\textbf{345.84±1.80}}  \\
                              & \multirow{2}{*}{U-rand}      & 3.5
                              & 362.58\textbf{\textbf{±}}0.93 & {\cellcolor[rgb]{0.753,0.753,0.753}}\textbf{\textbf{359.44±1.02}} & 359.94\textbf{\textbf{±}}0.64                                      \\
                               &                              & 4.0
                              & 366.46\textbf{\textbf{±}}2.44 & {\cellcolor[rgb]{0.753,0.753,0.753}}\textbf{\textbf{363.44±1.91}} & 374.26\textbf{\textbf{±}}2.71                                      \\
\bottomrule
\end{tabular}

\end{table}

Tables~\ref{table:noiseattack1} and~\ref{table:noiseattack_other} extend the same analysis to additional controllers and datasets. The gains persist for both traditional and RL-based controllers, showing that the defense is not tied to one specific policy architecture. The overall pattern is that MDRC retains its advantage across both stochastic and adaptive attacks and across both attack strengths evaluated, instead of failing once the attacker becomes policy-aware.

\begin{table*}[ht]
\caption{Performance of ATT in $JiNan_2$, $HangZhou_1$. Our MDRC recovers the state of traditional and RL-based TSC algorithms.}
\label{table:noiseattack_other}
\centering
   \fontsize{7}{8}\selectfont
\renewcommand{\arraystretch}{1.5}
\setlength{\tabcolsep}{1pt}
\begin{tabular}{ccccccccccccc} 
\toprule
\multirow{2}{*}{Dataset}      & \multirow{2}{*}{\begin{tabular}[c]{@{}c@{}}Noise \\Type\end{tabular}} & \multirow{2}{*}{\begin{tabular}[c]{@{}c@{}}~Noise \\Scale\end{tabular}} & FixedTime               & \multicolumn{3}{c}{Advanced-CoLight}                                                                                                                                 & \multicolumn{3}{c}{Advanced-MpLight}    & \multicolumn{3}{c}{Advanced-MaxPressure}                                                                                                      \\ 
\cline{4-13}
                              &                                                                       &                                                                         & base                    & base                            & RobustLight                                                       & MDRC                                                      & base                  & RobustLight                                              & MDRC      
                              & base                  
                              & RobustLight                                              
                              & MDRC                                        \\ 
\hline
\multirow{8}{*}{$JiNan_2$}    & \multirow{2}{*}{Gaussian}                                             & 3.5                                                                     & \multirow{8}{*}{368.77} & 338.12±12.82   & 292.70±2.31                                     & {\cellcolor[rgb]{0.753,0.753,0.753}}\textbf{\textbf{273.68±2.32}}  & 626.67±177.54  & 275.78±2.17                                     & {\cellcolor[rgb]{0.753,0.753,0.753}}\textbf{268.58±1.28}   & 276.10±1.50 & 280.43±1.77 & {\cellcolor[rgb]{0.753,0.753,0.753}}\textbf{269.98±0.81}\\
                              &                                                                       & 4.0                                                                     &                         & 357.56±30.63   & 308.17±3.60                                     & {\cellcolor[rgb]{0.753,0.753,0.753}}\textbf{\textbf{265.32±1.31}}  & 689.61±202.25  & 284.35±2.65                                     & {\cellcolor[rgb]{0.753,0.753,0.753}}\textbf{273.24±4.24}   & 281.06±1.67 & 287.71±3.12 & {\cellcolor[rgb]{0.753,0.753,0.753}}\textbf{274.22±1.07}\\
                              & \multirow{2}{*}{U-rand}                                               & 3.5                                                                     &                         & 748.76±97.15  & 564.09±61.36                                    & {\cellcolor[rgb]{0.753,0.753,0.753}}\textbf{\textbf{353.52±4.84}}  & 506.24±105.45 & 297.90±4.67 & {\cellcolor[rgb]{0.753,0.753,0.753}}\textbf{290.15±7.68}                                      & 301.54±2.38 & {\cellcolor[rgb]{0.753,0.753,0.753}}\textbf{300.41±1.79} & 304.85±4.04\\
                              &                                                                       & 4.0                                                                     &                         & 784.89±68.08  & 611.86±54.28                                    & {\cellcolor[rgb]{0.753,0.753,0.753}}\textbf{\textbf{334.67±4.53}}  & 572.98±126.74 & 303.68±5.53 & {\cellcolor[rgb]{0.753,0.753,0.753}}\textbf{295.55±8.74}                                      & 307.22±0.96 & {\cellcolor[rgb]{0.753,0.753,0.753}}\textbf{307.13±1.4} & 320.89±3.23\\
                              & \multirow{2}{*}{MAD}                                                  & 3.5                                                                     &                         & 563.99±40.13   & {\cellcolor[rgb]{0.753,0.753,0.753}}\textbf{273.66±4.84}                                     & 275.88±1.76  & 330.25±130.61 & {\cellcolor[rgb]{0.753,0.753,0.753}}\textbf{254.85±2.30}                                    & 262.66±0.89   & - & - & -\\
                              &                                                                       & 4.0                                                                     &                         & 630.31±19.84  & {\cellcolor[rgb]{0.753,0.753,0.753}}\textbf{277.99±1.23}                                     & 286.36±2.13  & 454.53±256.35 & {\cellcolor[rgb]{0.753,0.753,0.753}}\textbf{257.66±3.72}                                    & 279.84±0.91   & - & - & -\\
                              & \multirow{2}{*}{MinQ}                                                 & 3.5                                                                     &                         & 344.12±29.23  & {\cellcolor[rgb]{0.753,0.753,0.753}}\textbf{287.39±12.74}                                    & 292.35±8.27  & 272.16±8.66 & {\cellcolor[rgb]{0.753,0.753,0.753}}\textbf{256.80±2.05}                                    & 263.82±1.75   & - & - & -\\
                              &                                                                       & 4.0                                                                     &                         & 367.51±55.65 & 295.08±16.33                                    & {\cellcolor[rgb]{0.753,0.753,0.753}}\textbf{\textbf{288.83±4.36}}  & 402.47±141.56 & {\cellcolor[rgb]{0.753,0.753,0.753}}\textbf{261.35±3.43} &                                   277.89±2.16   & - & - & -\\ 
\hline
\multirow{8}{*}{$HangZhou_1$}    & \multirow{2}{*}{Gaussian}                                             & 3.5                                                                     & \multirow{8}{*}{495.57} & 512.63±25.89   & 363.83±8.25                                     & {\cellcolor[rgb]{0.753,0.753,0.753}}\textbf{337.84±2.34}  & 334.03±2.48 & 351.44±9.07                                     & {\cellcolor[rgb]{0.753,0.753,0.753}}\textbf{315.75±1.77}   & 327.94±1.03 & 346.14±1.58 & {\cellcolor[rgb]{0.753,0.753,0.753}}\textbf{319.10±0.83}\\
                              &                                                                       & 4.0                                                                     &                         & 530.59±30.78  & 388.49±7.73                                     & {\cellcolor[rgb]{0.753,0.753,0.753}}\textbf{\textbf{336.53±2.89}}  & 338.76±3.89 & 381.53±13.58                                     & {\cellcolor[rgb]{0.753,0.753,0.753}}\textbf{316.92±1.54}   & 331.25±1.28 & 370.61±2.56 & {\cellcolor[rgb]{0.753,0.753,0.753}}\textbf{320.27±0.43}\\
                              & \multirow{2}{*}{U-rand}                                               & 3.5                                                                     &                         & 971.03±34.19  & 537.52±56.18                                    & {\cellcolor[rgb]{0.753,0.753,0.753}}\textbf{\textbf{426.79±23.70}}  & 354.63±2.05 & {\cellcolor[rgb]{0.753,0.753,0.753}}\textbf{332.80±1.60}                                    & 343.87±5.61   & 355.51±3.32 & {\cellcolor[rgb]{0.753,0.753,0.753}}\textbf{348.57±1.07} & 350.61±1.76 \\
                              &                                                                       & 4.0                                                                     &                         & 982.41±42.18  & 561.60±50.29                                    & {\cellcolor[rgb]{0.753,0.753,0.753}}\textbf{\textbf{408.59±14.11}}  & 360.11±2.75 & {\cellcolor[rgb]{0.753,0.753,0.753}}\textbf{334.80±1.67}                                    & 351.45±3.08   & 360.88±3.08 & {\cellcolor[rgb]{0.753,0.753,0.753}}\textbf{353.11±1.71} & 363.21±4.21\\
                              & \multirow{2}{*}{MAD}                                                  & 3.5                                                                     &                         & 751.58±44.00  & {\cellcolor[rgb]{0.753,0.753,0.753}}\textbf{319.42±1.81} & 333.67±3.23                                      & 308.78±3.91 & 312.76±4.85                                    & {\cellcolor[rgb]{0.753,0.753,0.753}}\textbf{309.87±2.10}   & - & - & -\\
                              &                                                                       & 4.0                                                                     &                         & 754.66±37.04  & {\cellcolor[rgb]{0.753,0.753,0.753}}\textbf{\textbf{323.48±2.45}} & 352.27±4.76                                      & 313.53±3.06 & {\cellcolor[rgb]{0.753,0.753,0.753}}\textbf{318.14±5.43}                                    & 331.27±4.13   & - & - & -\\
                              & \multirow{2}{*}{MinQ}                                                 & 3.5                                                                     &                         & 506.60±42.19  & {\cellcolor[rgb]{0.753,0.753,0.753}}\textbf{35.29±3.50}                                    & 351.39±11.40  & 320.32±6.34 & {\cellcolor[rgb]{0.753,0.753,0.753}}\textbf{314.07±4.20}                                    & 315.70±3.19  & - & - & -\\
                              &                                                                       & 4.0                                                                     &                         & 550.00±37.44   & {\cellcolor[rgb]{0.753,0.753,0.753}}\textbf{343.82±5.70}                                    & 363.18±14.05  & 324.65±5.88 & {\cellcolor[rgb]{0.753,0.753,0.753}}\textbf{323.04±4.15}                                    & 330.05±5.85   & - & - & -\\ 
\bottomrule
\end{tabular}

\end{table*}

\begin{table*}[ht]
\caption{ATT in JiNan and HangZhou: 25\% refers to missing data in $sensor_W$, and 50\% refers to $sensor_W$ and $sensor_E$.}
\label{table:unmask}
\centering
\footnotesize
\fontsize{7}{8}\selectfont
\renewcommand{\arraystretch}{1.5}
\setlength{\tabcolsep}{2.5pt}

\begin{tabular}{ccccccccc} 
\toprule
\multirow{2}{*}{Dataset}   & \multirow{2}{*}{Mask Scale} & FixedTime                    & \multicolumn{3}{c}{Advanced-MaxPressure}                                                                                                      & \multicolumn{3}{c}{Advanced-MpLight}                                                                                                            \\ 
\cline{3-9}
                           &                             & base                         & base                  & RobustLight                                               & MDRC                                             & base                   & RobustLight                                               & MDRC                                              \\ 
\hline
\multirow{2}{*}{$JiNan_1$}    & 25\%                        & \multirow{2}{*}{428.11±0.00} & 352.13\textbf{±}0.00  & {\cellcolor[rgb]{0.753,0.753,0.753}}\textbf{296.50±1.12}  & 315.25\textbf{±}6.42                                      & 552.15\textbf{±}120.94  & {\cellcolor[rgb]{0.753,0.753,0.753}}\textbf{371.95±90.21} & 400.35\textbf{±}63.58                                      \\
                           & 50\%                        &                              & 1059.67\textbf{±}0.00 & 610.43\textbf{±}68.52                                     & {\cellcolor[rgb]{0.753,0.753,0.753}}\textbf{432.15±4.95} & 1045.75\textbf{±}26.83  & 878.09\textbf{±}16.55 & {\cellcolor[rgb]{0.753,0.753,0.753}}\textbf{867.15±60.67}                                      \\ 
\hline
\multirow{2}{*}{$JiNan_2$}    & 25\%                        & \multirow{2}{*}{368.76±0.00} & 323.13\textbf{±}0.00  & {\cellcolor[rgb]{0.753,0.753,0.753}}\textbf{273.18±3.95}  & 278.87\textbf{±}3.68                                      & 490.56\textbf{±}92.29  & {\cellcolor[rgb]{0.753,0.753,0.753}}\textbf{276.05±4.71}  & 312.99\textbf{±}47.86                                      \\
                           & 50\%                        &                              & 1209.97\textbf{±}0.00 & {\cellcolor[rgb]{0.753,0.753,0.753}}\textbf{755.51±106.57} & 912.18\textbf{±}24.89                                     & 1082.64\textbf{±}65.15 & 612.14\textbf{±}43.37                                     & {\cellcolor[rgb]{0.753,0.753,0.753}}\textbf{558.87±71.86}  \\ 
\hline
\multirow{2}{*}{$JiNan_3$}    & 25\%                        & \multirow{2}{*}{383.01±0.00} & 340.81\textbf{±}0.00  & 281.56\textbf{±}4.11   & {\cellcolor[rgb]{0.753,0.753,0.753}}\textbf{281.05±4.23}                                      & 403.29\textbf{±}30.61   & {\cellcolor[rgb]{0.753,0.753,0.753}}\textbf{288.68±9.41} & 337.98\textbf{±}19.97                                       \\
                           & 50\%                        &                              & 1109.57\textbf{±}0.00 & 570.54\textbf{±}50.20                                      & {\cellcolor[rgb]{0.753,0.753,0.753}}\textbf{371.44±21.02}  & 1061.35\textbf{±}67.51 & 918.43\textbf{±}27.84                                     & {\cellcolor[rgb]{0.753,0.753,0.753}}\textbf{654.51±83.29}  \\ 
\hline
\multirow{2}{*}{$HangZhou_1$} & 25\%                        & \multirow{2}{*}{495.57±0.00} & 530.33\textbf{±}0.00  & 369.52\textbf{±}12.54                                     & {\cellcolor[rgb]{0.753,0.753,0.753}}\textbf{318.41±5.98}  & 478.89\textbf{±}37.35    & 363.55\textbf{±}7.80                                    & {\cellcolor[rgb]{0.753,0.753,0.753}}\textbf{343.70±10.98}   \\
                           & 50\%                        &                              & 1186.56\textbf{±}0.00 & 563.56\textbf{±}36.21                                     & {\cellcolor[rgb]{0.753,0.753,0.753}}\textbf{440.00±94.94}  & 867.95\textbf{±}172.63 & 824.83\textbf{±}149.15                                    & {\cellcolor[rgb]{0.753,0.753,0.753}}\textbf{394.88±43.05}  \\ 
\hline
\multirow{2}{*}{$HangZhou_2$} & 25\%                        & \multirow{2}{*}{406.65±0.00} & 409.56\textbf{±}0.00  & 350.86\textbf{±}3.11                                      & {\cellcolor[rgb]{0.753,0.753,0.753}}\textbf{341.47±1.68}  & 373.59\textbf{±}13.70  & 360.22\textbf{±}10.27                                      & {\cellcolor[rgb]{0.753,0.753,0.753}}\textbf{346.86±6.61}   \\
                           & 50\%                        &                              & 782.93\textbf{±}0.00  & 447.28\textbf{±}15.87                                      & {\cellcolor[rgb]{0.753,0.753,0.753}}\textbf{350.47±2.12}  & 633.73\textbf{±}89.70  & 459.36\textbf{±}6.87                                      & {\cellcolor[rgb]{0.753,0.753,0.753}}\textbf{355.15±7.12}   \\
\bottomrule
\end{tabular}

\end{table*}

\takeawaybox{RQ1}{Across stochastic and adaptive observation attacks, MDRC lowers ATT on average and provides substantial gains in many high-corruption settings, although improvements are not universal across all controller–attack pairs.}

\noindent\textbf{RQ2: Can MDRC maintain control quality under sensor failures and structured missingness?}
\label{subsubsec:sensor_damage_state}

Table~\ref{table:unmask} evaluates availability attacks by masking sensor inputs. Here, 25\% masking corresponds to the failure of $sensor_W$, while 50\% masking corresponds to the joint failure of $sensor_W$ and $sensor_E$. This is a harder setting than small bounded perturbations because the controller must operate under partial observability. MDRC uses the Repaint recovery procedure to reconstruct missing state entries and achieves an average improvement of 12.75\% over RobustLight. In the worst setting reported here, $HangZhou_1$ with 50\% sensor damage, the gain reaches 52.13\%. Notably, even when standard controllers degrade sharply under 50\% masking, MDRC often keeps them competitive with or better than FixedTime, indicating that the defense remains effective under severe sensing failures.

Table~\ref{table:unmask1} provides controller-specific evidence on Advanced-CoLight and shows the same trend: MDRC is especially valuable in the more degraded 50\% masking regime. Together, these results indicate that MDRC improves worst-case behavior under structured missingness rather than only helping in mild failure cases.

\begin{table*}[ht]
\centering
   \fontsize{7}{8}\selectfont
\renewcommand{\arraystretch}{1.5}
\setlength{\tabcolsep}{0.8pt}
\caption{Transfer to unseen cities performance of ATT in $JiNan$, $HangZhou$. Results are
averaged over 5 runs.}
\label{table:transfer_zero_few_shot}
\begin{tabular}{cccccccccccc} 
\toprule
\multirow{2}{*}{Dataset}      & \multirow{2}{*}{\begin{tabular}[c]{@{}c@{}}Noise \\Type\end{tabular}} & \multirow{2}{*}{\begin{tabular}[c]{@{}c@{}}~Noise \\Scale\end{tabular}} & \multirow{2}{*}{FixedTime} & \multicolumn{4}{c}{Advanced-CoLight}                                                                                                                                                                  & \multicolumn{4}{c}{Advanced-MpLight}                                                                                                                                                                   \\ 
\cline{5-12}
                              &                                                                       &                                                                         &                            & base   & RobustLight                                         & \begin{tabular}[c]{@{}c@{}}MDRC\\(Zero-Shot)\end{tabular} & \begin{tabular}[c]{@{}c@{}}MDRC\\(Few-Shot)\end{tabular} & base   & RobustLight                                         & \begin{tabular}[c]{@{}c@{}}MDRC\\(Zero-Shot)\end{tabular} & \begin{tabular}[c]{@{}c@{}}MDRC\\(Few-Shot)\end{tabular}  \\ 
\hline
\multirow{5}{*}{$JiNan_1$}    & Gaussian                                                              & 3.5                                                                     & \multirow{5}{*}{428.11}    & 316.96 & 423.56                                              & 329.56                                                             & {\cellcolor[rgb]{0.753,0.753,0.753}}\textbf{285.88}               & 327.93 & {\cellcolor[rgb]{0.753,0.753,0.753}}\textbf{322.04} & 380.98                                                             & 369.81                                                             \\
                              & U-rand                                                                & 3.5                                                                     &                            & 483.26 & 540.52                                              & 419.14                                                             & {\cellcolor[rgb]{0.753,0.753,0.753}}\textbf{360.59}               & 417.81 & 328.01                                              & 340.18                                                             & {\cellcolor[rgb]{0.753,0.753,0.753}}\textbf{325.05}                \\
                              & MAD                                                                   & 3.5                                                                     &                            & 454.77 & 483.18                                              & 445.61                                                             & {\cellcolor[rgb]{0.753,0.753,0.753}}\textbf{325.73}               & 339.41 & {\cellcolor[rgb]{0.753,0.753,0.753}}\textbf{296.69} & 495.94                                                             & 454.51                                                             \\
                              & MinQ                                                                  & 3.5                                                                     &                            & 493.10 & 393.85                                              & 528.60                                                             & {\cellcolor[rgb]{0.753,0.753,0.753}}\textbf{347.36}               & 347.62 & 368.18                                              & 395.84                                                             & {\cellcolor[rgb]{0.753,0.753,0.753}}\textbf{339.12}                \\
                              & Mask~                                                                 & 25\%                                                                    &                            & 343.34 & 336.82                                              & 291.41                                                             & {\cellcolor[rgb]{0.753,0.753,0.753}}\textbf{312.83}               & 552.15 & {\cellcolor[rgb]{0.753,0.753,0.753}}\textbf{398.48} & 562.94                                                             & 639.32                                                             \\ 
\hline
\multirow{5}{*}{$JiNan_2$}    & Gaussian                                                              & 3.5                                                                     & \multirow{5}{*}{368.77}    & 338.12 & 427.64                                              & 292.83                                                             & {\cellcolor[rgb]{0.753,0.753,0.753}}\textbf{268.72}               & 626.67 & 291.99                                              & {\cellcolor[rgb]{0.753,0.753,0.753}}\textbf{285.45}                & 287.36                                                             \\
                              & U-rand                                                                & 3.5                                                                     &                            & 748.76 & 565.60                                              & 619.69                                                             & {\cellcolor[rgb]{0.753,0.753,0.753}}\textbf{314.64}               & 506.24 & {\cellcolor[rgb]{0.753,0.753,0.753}}\textbf{308.08} & 362.71                                                             & 333.73                                                             \\
                              & MAD                                                                   & 3.5                                                                     &                            & 563.99 & 356.23                                              & 487.48                                                             & {\cellcolor[rgb]{0.753,0.753,0.753}}\textbf{273.49}               & 330.25 & 280.68                                              & 270.75                                                             & {\cellcolor[rgb]{0.753,0.753,0.753}}\textbf{268.21}                \\
                              & MinQ                                                                  & 3.5                                                                     &                            & 344.12 & 348.00                                              & 368.26                                                             & {\cellcolor[rgb]{0.753,0.753,0.753}}\textbf{266.12}               & 272.16 & 271.29                                              & 277.54                                                             & {\cellcolor[rgb]{0.753,0.753,0.753}}\textbf{270.85}                \\
                              & Mask~                                                                 & 25\%                                                                    &                            & 277.85 & 297.95                                              & 270.94                                                             & {\cellcolor[rgb]{0.753,0.753,0.753}}\textbf{285.39}               & 490.56 & {\cellcolor[rgb]{0.753,0.753,0.753}}\textbf{342.39} & 359.96                                                             & 386.45                                                             \\ 
\hline
\multirow{5}{*}{$JiNan_3$}    & Gaussian                                                              & 3.5                                                                     & \multirow{5}{*}{383.01}    & 320.04 & 413.53                                              & 288.25                                                             & {\cellcolor[rgb]{0.753,0.753,0.753}}\textbf{262.37}               & 378.13 & 293.93                                              & 351.54                                                             & {\cellcolor[rgb]{0.753,0.753,0.753}}\textbf{283.70}                \\
                              & U-rand                                                                & 3.5                                                                     &                            & 481.55 & 543.19                                              & 639.80                                                             & {\cellcolor[rgb]{0.753,0.753,0.753}}\textbf{312.58}               & 496.63 & {\cellcolor[rgb]{0.753,0.753,0.753}}\textbf{307.64} & 729.41                                                             & 337.57                                                             \\
                              & MAD                                                                   & 3.5                                                                     &                            & 446.89 & 401.58                                              & 335.38                                                             & {\cellcolor[rgb]{0.753,0.753,0.753}}\textbf{260.04}               & 442.19 & 272.86                                              & 268.99                                                             & {\cellcolor[rgb]{0.753,0.753,0.753}}\textbf{263.00}                \\
                              & MinQ                                                                  & 3.5                                                                     &                            & 376.00 & 364.33                                              & 318.01                                                             & {\cellcolor[rgb]{0.753,0.753,0.753}}\textbf{262.94}               & 423.76 & 268.08                                              & 370.69                                                             & {\cellcolor[rgb]{0.753,0.753,0.753}}\textbf{261.61}                \\
                              & Mask~                                                                 & 25\%                                                                    &                            & 324.42 & 309.56                                              & 304.23                                                             & {\cellcolor[rgb]{0.753,0.753,0.753}}\textbf{287.33}               & 403.29 & 360.49                                              & 588.86                                                             & {\cellcolor[rgb]{0.753,0.753,0.753}}\textbf{335.52}                \\ 
\hline
\multirow{5}{*}{$HangZhou_1$} & Gaussian                                                              & 3.5                                                                     & \multirow{5}{*}{495.57}    & 512.63 & 442.47                                              & 683.91                                                             & {\cellcolor[rgb]{0.753,0.753,0.753}}\textbf{354.15}               & 334.03 & 355.64                                              & 336.00                                                             & {\cellcolor[rgb]{0.753,0.753,0.753}}\textbf{327.65}                \\
                              & U-rand                                                                & 3.5                                                                     &                            & 971.03 & 775.62                                              & 985.47                                                             & {\cellcolor[rgb]{0.753,0.753,0.753}}\textbf{615.19}               & 354.63 & 363.64                                              & 366.68                                                             & {\cellcolor[rgb]{0.753,0.753,0.753}}\textbf{353.87}                \\
                              & MAD                                                                   & 3.5                                                                     &                            & 751.58 & {\cellcolor[rgb]{0.753,0.753,0.753}}\textbf{513.22} & 860.74                                                             & 719.18                                                            & 308.78 & 547.87                                              & {\cellcolor[rgb]{0.753,0.753,0.753}}\textbf{312.55}                & 318.79                                                             \\
                              & MinQ                                                                  & 3.5                                                                     &                            & 506.60 & 467.94                                              & 678.96                                                             & {\cellcolor[rgb]{0.753,0.753,0.753}}\textbf{369.34}               & 320.32 & 507.84                                              & {\cellcolor[rgb]{0.753,0.753,0.753}}\textbf{306.32}                & 317.45                                                             \\
                              & Mask~                                                                 & 25\%                                                                    &                            & 418.49 & 334.30                                              & {\cellcolor[rgb]{0.753,0.753,0.753}}\textbf{324.31}                & 375.17                                                            & 478.89 & 442.75                                              & 456.33                                                             & {\cellcolor[rgb]{0.753,0.753,0.753}}\textbf{371.35}                \\ 
\hline
\multirow{5}{*}{$HangZhou_2$} & Gaussian                                                              & 3.5                                                                     & \multirow{5}{*}{406.65}    & 495.92 & 419.85                                              & 377.31                                                             & {\cellcolor[rgb]{0.753,0.753,0.753}}\textbf{343.34}               & 429.53 & 377.48                                              & 345.56                                                             & {\cellcolor[rgb]{0.753,0.753,0.753}}\textbf{335.54}                \\
                              & U-rand                                                                & 3.5                                                                     &                            & 567.56 & 585.79                                              & 568.48                                                             & {\cellcolor[rgb]{0.753,0.753,0.753}}\textbf{378.70}               & 481.32 & {\cellcolor[rgb]{0.753,0.753,0.753}}\textbf{365.99} & 390.19                                                             & 378.26                                                             \\
                              & MAD                                                                   & 3.5                                                                     &                            & 496.73 & 481.36                                              & 369.08                                                             & {\cellcolor[rgb]{0.753,0.753,0.753}}\textbf{357.09}               & 433.46 & 492.92                                              & 322.85                                                             & {\cellcolor[rgb]{0.753,0.753,0.753}}\textbf{321.66}                \\
                              & MinQ                                                                  & 3.5                                                                     &                            & 441.72 & 464.05                                              & 368.53                                                             & {\cellcolor[rgb]{0.753,0.753,0.753}}\textbf{344.55}               & 425.09 & 444.09                                              & {\cellcolor[rgb]{0.753,0.753,0.753}}\textbf{317.63}                & 328.69                                                             \\
                              & Mask~                                                                 & 25\%                                                                    &                            & 348.21 & 359.46                                              & 374.68                                                             & {\cellcolor[rgb]{0.753,0.753,0.753}}\textbf{340.42}               & 373.59 & {\cellcolor[rgb]{0.753,0.753,0.753}}\textbf{347.58} & 372.22                                                             & 361.43                                                             \\
\bottomrule
\end{tabular}

\end{table*}

\begin{table}[ht]
\caption{ATT in JiNan and HangZhou: 25\% refers to missing data in $sensor_W$, and 50\% refers to $sensor_W$ and $sensor_E$.}
\label{table:unmask1}
\footnotesize
\centering
   \fontsize{7}{8}\selectfont
\renewcommand{\arraystretch}{1.5}
\setlength{\tabcolsep}{4.8pt}
\begin{tabular}{ccccc} 
\toprule
\multirow{2}{*}{Dataset}   & \multirow{2}{*}{Mask Scale}   & \multicolumn{3}{c}{Advanced-CoLight}                                                                       \\ 
\cline{3-5}
                           &   & base                  & RobustLight           & MDRC                                              \\ 
\hline
\multirow{2}{*}{JiNan1}    & 25\%  & 343.34\textbf{±}8.39  & 310.14\textbf{±}9.63 & {\cellcolor[rgb]{0.753,0.753,0.753}}\textbf{307.98±7.15}   \\
                           & 50\%
                           & 699.02\textbf{±}35.07 & {\cellcolor[rgb]{0.753,0.753,0.753}}\textbf{539.30±54.88} & 648.73\textbf{±}20.38   \\ 
\hline
\multirow{2}{*}{JiNan2}    & 25\%  & 277.85\textbf{±}7.50  & {\cellcolor[rgb]{0.753,0.753,0.753}}\textbf{266.62±3.79}  & 270.90\textbf{±}6.33   \\
                           & 50\%
                           & 682.61\textbf{±}29.58 & 351.08\textbf{±}17.45 & {\cellcolor[rgb]{0.753,0.753,0.753}}\textbf{322.20±8.94}   \\ 
\hline
\multirow{2}{*}{JiNan3}    & 25\%  & 324.42\textbf{±}13.55 & 278.06\textbf{±}9.84 & {\cellcolor[rgb]{0.753,0.753,0.753}}\textbf{267.95±6.13}   \\
                           & 50\%
                           & 627.95\textbf{±}53.19 & 417.49\textbf{±}111.67 & {\cellcolor[rgb]{0.753,0.753,0.753}}\textbf{306.57±7.86}  \\ 
\hline
\multirow{2}{*}{HangZhou1} & 25\%  & 418.49\textbf{±}13.59 & {\cellcolor[rgb]{0.753,0.753,0.753}}\textbf{331.65±7.68}  & 383.35\textbf{±}12.47   \\
                           & 50\%
                           & 624.83\textbf{±}13.27 & 434.31\textbf{±}34.36 & {\cellcolor[rgb]{0.753,0.753,0.753}}\textbf{415.23±7.23}   \\ 
\hline
\multirow{2}{*}{HangZhou2} & 25\%  & 348.21\textbf{±}7.21   & 340.80\textbf{±}4.64  & {\cellcolor[rgb]{0.753,0.753,0.753}}\textbf{335.02±2.35}   \\
                           & 50\%
                           & 499.34\textbf{±}28.56 & 453.90\textbf{±}26.08 & {\cellcolor[rgb]{0.753,0.753,0.753}}\textbf{352.05±6.22}   \\
\bottomrule
\end{tabular}

\end{table}

\takeawaybox{RQ2}{MDRC remains effective under partial observability and is most useful in the harder masking regime. It improves ATT by 12.75\% on average over RobustLight and yields up to 52.13\% gain under 50\% sensor damage.}

\noindent\textbf{RQ3: Does MDRC transfer to unseen cities under attack in zero-shot and few-shot settings?}

Table~\ref{table:attack_sumo} evaluates transfer to the unseen SUMO Cologne8 dataset. The diffusion outer learner is meta-trained on Cityflow datasets from JiNan and HangZhou, then deployed zero-shot and adapted few-shot with 50 samples on Cologne8. Under Gaussian and U-rand attacks, the few-shot setting consistently improves over zero-shot for both emergency and regular vehicles, showing that MDRC can adapt to a new city with limited target data while preserving robustness under attack.

\begin{table}[ht]
\centering
\fontsize{7}{8}\selectfont
\renewcommand{\arraystretch}{1.2}
\setlength{\tabcolsep}{2.5pt}
\caption{Transfer performance (AWT and ATT) of EMV and REV on Cologne8. Results are averaged over 5 runs.}
\label{table:attack_sumo}

\begin{tabular}{c c c ccc ccc}
\toprule
Algorithm & Metrics &
\makecell{Noise-\\Free} &
Gaussian &
\makecell{Zero-\\shot} &
\makecell{Few-\\shot} &
U-Rand &
\makecell{Zero-\\shot} &
\makecell{Few-\\shot} \\
\midrule
\multirow{4}{*}{MPlight}
 & $AWT_{EMV}$ & 8 & 25.0 & 20.0 & \cellcolor[rgb]{0.753,0.753,0.753}\textbf{15.0} & 20.0 & 15.0 & \cellcolor[rgb]{0.753,0.753,0.753}\textbf{15.0} \\
 & $ATT_{EMV}$ & 20 & 25.0 & 20.0 & \cellcolor[rgb]{0.753,0.753,0.753}\textbf{15.0} & 20.0 & 15.0 & \cellcolor[rgb]{0.753,0.753,0.753}\textbf{15.0} \\
 & $AWT_{REV}$ & 0.51 & 2.14 & \cellcolor[rgb]{0.753,0.753,0.753}\textbf{1.85} & 1.95 & 2.37 & 2.16 & \cellcolor[rgb]{0.753,0.753,0.753}\textbf{2.13} \\
 & $ATT_{RMV}$ & 54.40 & 67.47 & 62.50 & \cellcolor[rgb]{0.753,0.753,0.753}\textbf{62.17} & 67.49 & 65.57 & \cellcolor[rgb]{0.753,0.753,0.753}\textbf{64.96} \\
\midrule
\multirow{4}{*}{MaxPressure}
 & $AWT_{EMV}$ & 6.0 & 0.0 & 0.0 & \cellcolor[rgb]{0.753,0.753,0.753}\textbf{0.0} & 0.0 & \cellcolor[rgb]{0.753,0.753,0.753}\textbf{0.0} & 6.0 \\
 & $ATT_{EMV}$ & 25.0 & 15.0 & 15.0 & \cellcolor[rgb]{0.753,0.753,0.753}\textbf{15.0} & 15.0 & \cellcolor[rgb]{0.753,0.753,0.753}\textbf{15.0} & 20.0 \\
 & $AWT_{REV}$ & 0.68 & 2.25 & 1.48 & \cellcolor[rgb]{0.753,0.753,0.753}\textbf{1.43} & 3.32 & \cellcolor[rgb]{0.753,0.753,0.753}\textbf{2.21} & 2.24 \\
 & $ATT_{REV}$ & 56.46 & 66.71 & 61.64 & \cellcolor[rgb]{0.753,0.753,0.753}\textbf{61.29} & 72.97 & 65.39 & \cellcolor[rgb]{0.753,0.753,0.753}\textbf{64.52} \\
\bottomrule
\end{tabular}
\end{table}

\begin{figure*}[ht]
    \centering
    \includegraphics[width=1\linewidth]{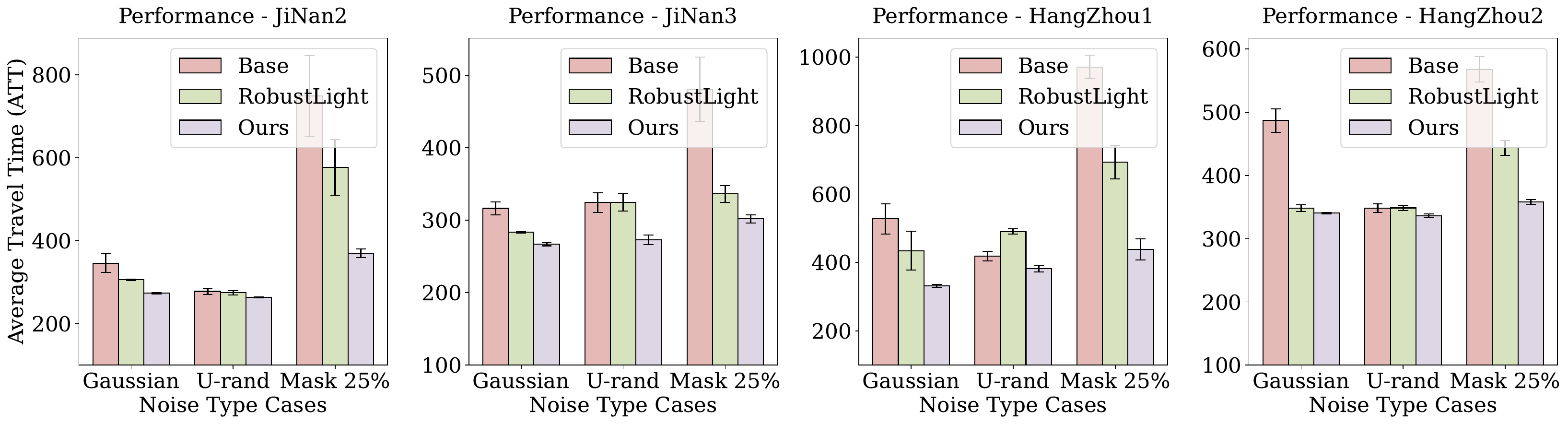}
    \caption{Performance of MDRC based on Advanced-Colight in $JiNan_2$, $JiNan_3$, $HangZhou_1$, and $HangZhou_2$ transferred from $JiNan_1$ (noise scale is 3.5).}
    \label{fig:transfer_other}
\end{figure*}

\begin{figure}[ht]
    \centering
    \includegraphics[width=1\linewidth]{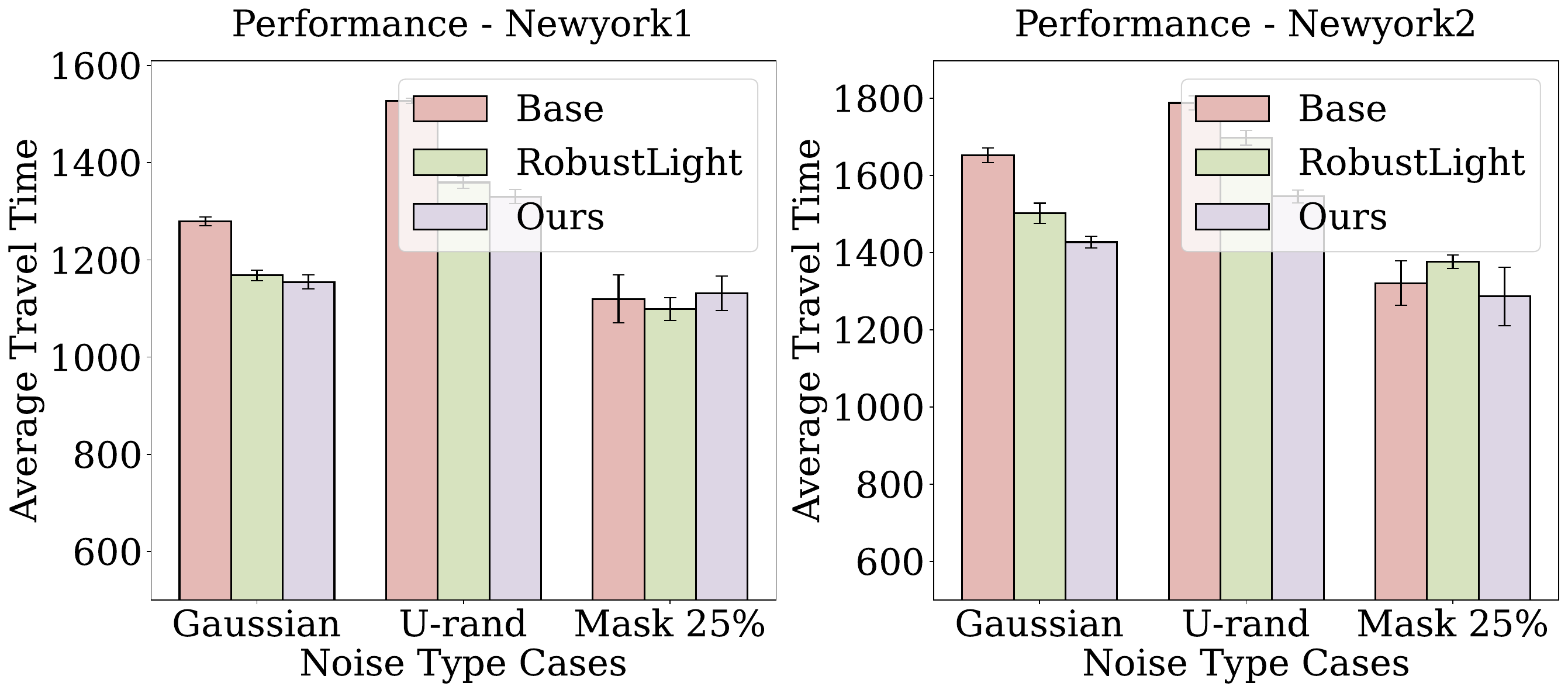}
    \caption{Zero-shot transfer study in $Newyork_1$ and $Newyork_2$.}
    \label{fig:transfer}
\end{figure}

Table~\ref{table:transfer_zero_few_shot} studies cross-city transfer between JiNan and HangZhou under multiple attack types. MDRC (Zero-Shot) already outperforms RobustLight in many heavy-noise settings, especially under MAD and Mask, while MDRC (Few-Shot) with only 100 target samples achieves the best overall performance and often recovers or surpasses clean-data performance. This addresses a core deployment question for security defenses: whether robustness survives domain shift instead of being restricted to the training city.

We also test larger-scale transfer from JiNan and HangZhou to NewYork. Figure~\ref{fig:transfer} shows that MDRC improves zero-shot transfer performance by 6.92\% on the NewYork setting, and Figure~\ref{fig:transfer_other} extends the study to six datasets under Gaussian, U-rand, and Mask perturbations. MDRC improves most transferred settings, including $Newyork_1$ Gaussian and $Newyork_2$ Gaussian. At the same time, Figure~\ref{fig:transfer_other} reveals a meaningful failure case: on $Newyork_1$ under Mask, MDRC underperforms RobustLight. This suggests that transfer under severe structured missingness can still depend on target-domain similarity, which helps delineate the current boundary of the defense.

\takeawaybox{RQ3}{Few-shot adaptation generally strengthens transfer, particularly under stochastic and adaptive perturbations, although gains are not universal under structured.}

\noindent\textbf{RQ4: Which components of MDRC are necessary for the robustness-latency trade-off?}

\begin{figure*}[ht]
    \centering
    \begin{subfigure}[t]{0.45\textwidth}
        \centering
        \includegraphics[width=\textwidth]{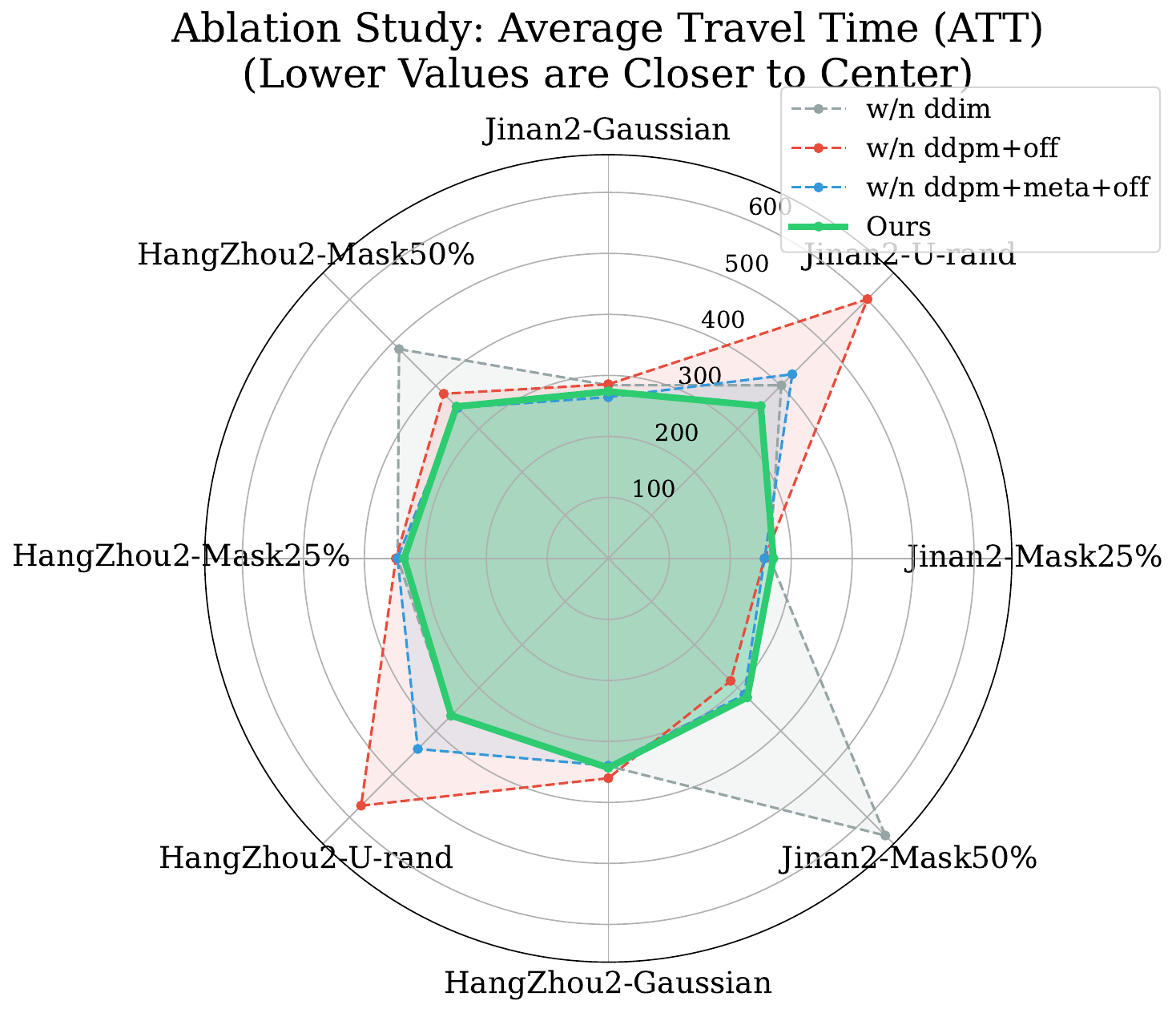}
        \caption{Average Travel Time.}
        \label{fig:ablation_att}
    \end{subfigure}
    \hfill
    \begin{subfigure}[t]{0.45\textwidth}
        \centering
        \includegraphics[width=\textwidth]{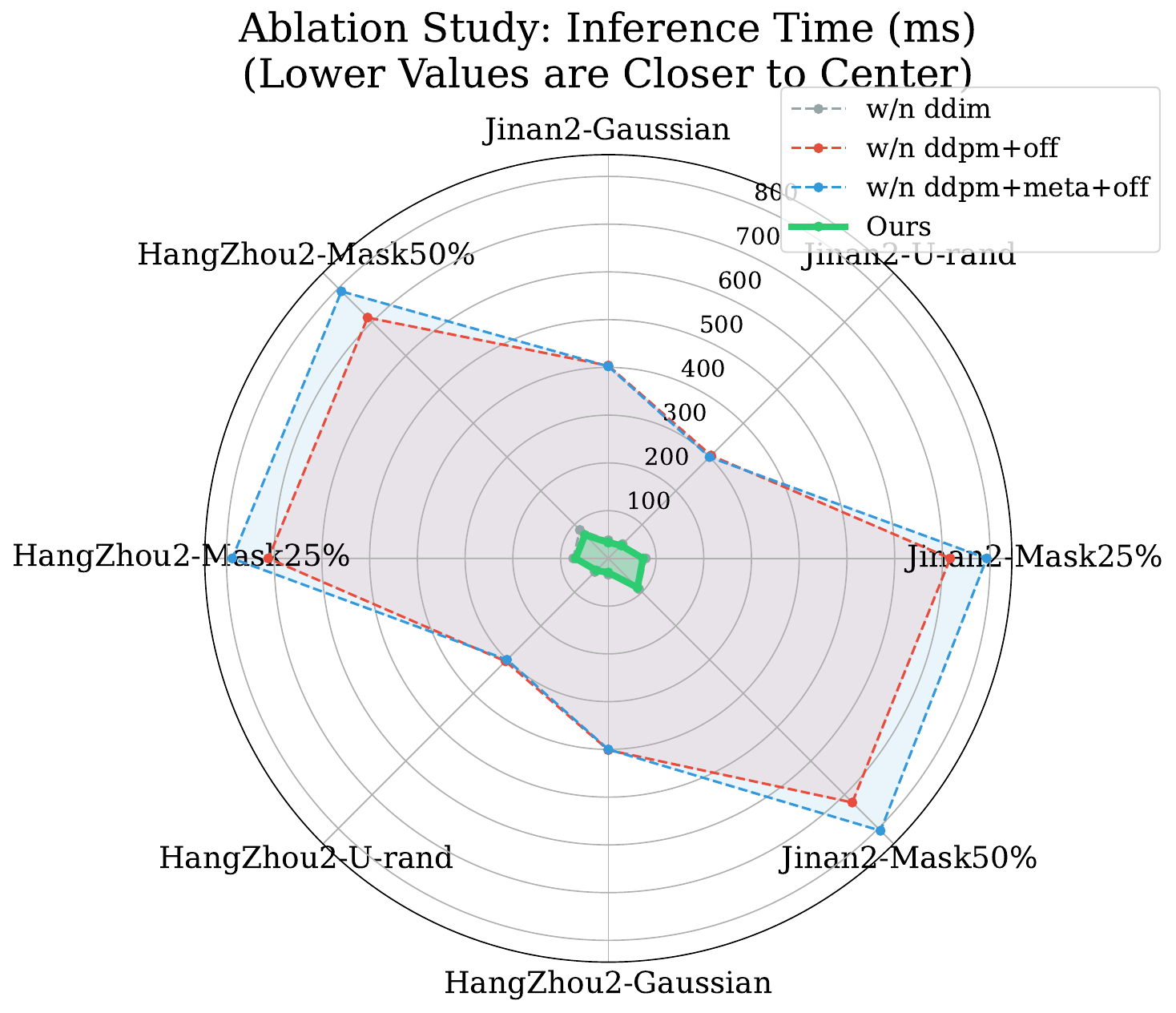}
        \caption{Inference time.}
        \label{fig:ablation_inference}
    \end{subfigure}
    \caption{Ablation studies on ATT and inference time in $HangZhou_2$ and $JiNan_2$.}
    \label{fig:ablation combined}
\end{figure*}

\begin{figure}[ht]
    \centering
    \includegraphics[width=1\linewidth]{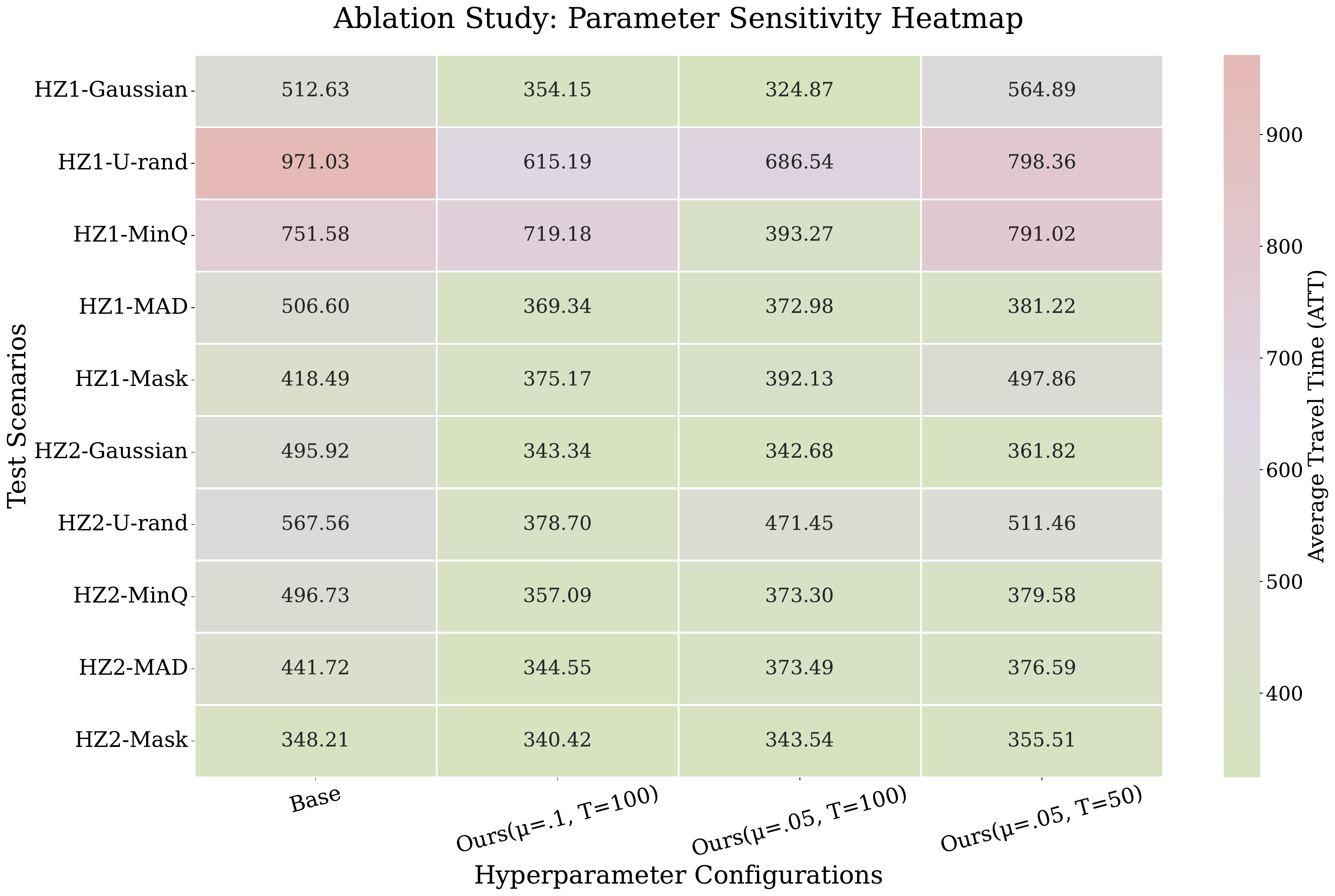}
    \caption{Ablation study of few-shot (100 samples) transfer performance from $JiNan$ to $HangZhou$ under different noise types and scales based on Advanced-CoLight with different learning rates and diffusion steps.}
    \label{fig:abalation_lr_T}
\end{figure}

We compare four variants: DDIM with online training, DDPM with offline training (DMBP~\citep{yang2023dmbp}), MDRC with DDPM, and full MDRC. Figure~\ref{fig:ablation combined} shows two consistent patterns. First, combining meta-learning with diffusion is important for robustness: MDRC achieves the lowest ATT across most noise and mask settings. Second, using DDIM is important for efficiency: the full MDRC model is substantially faster than DDPM-based alternatives. Figure~\ref{fig:abalation_lr_T} further shows that, in few-shot transfer from $JiNan$ to $HangZhou$, a meta learning rate of 0.1 and 100 diffusion steps yield the best overall trade-off. These ablations indicate that the security benefit is not attributable to a single implementation trick; both the meta-learning component and the fast diffusion sampler matter.

\takeawaybox{RQ4}{The robustness-latency trade-off depends on both ingredients of MDRC: meta-learning improves recovery quality, and DDIM keeps inference fast. Among the tested settings, a meta learning rate of 0.1 with 100 diffusion steps gives the best few-shot transfer trade-off.}

\noindent\textbf{RQ5: Is MDRC practical as an online plug-in defense?}

Compared to RobustLight, MDRC reduces denoising runtime by 87.9\% and demasking runtime by 91.9\%, as shown in Table~\ref{table:directmetric}. These reductions are directly relevant to deployment because MDRC is intended to run between sensing and control in the online loop.

\begin{table}[ht]
\caption{Inference time comparison (in milliseconds) based on Advanced-Colight.}
\label{table:directmetric}
\small
\centering
   \fontsize{7}{8}\selectfont
\renewcommand{\arraystretch}{1.2}
\setlength{\tabcolsep}{5.5pt}

\begin{tabular}{cccccc}
    \toprule
    \multicolumn{3}{c}{\textbf{$JiNan_1$}} & \multicolumn{3}{c}{\textbf{$HangZhou_1$}} \\ 
    \cmidrule(lr){1-3} \cmidrule(lr){4-6}
    Type & RobustLight & \textbf{Our} & Type & RobustLight & \textbf{Our} \\ 
    \midrule
    Gaussian 
        & 131.52  
        & {\cellcolor[rgb]{0.85,0.85,0.85}}\textbf{33.40} 
        & Gaussian 
        & 139.60  
        & {\cellcolor[rgb]{0.85,0.85,0.85}}\textbf{31.89} \\

    U-rand 
        & 173.76  
        & {\cellcolor[rgb]{0.85,0.85,0.85}}\textbf{38.23} 
        & U-rand 
        & 179.18
        & {\cellcolor[rgb]{0.85,0.85,0.85}}\textbf{36.24} \\

    MAD 
        & 1049.95 
        & {\cellcolor[rgb]{0.85,0.85,0.85}}\textbf{119.19} 
        & MAD 
        & 1505.61
        & {\cellcolor[rgb]{0.85,0.85,0.85}}\textbf{162.36} \\

    MinQ 
        & 1081.03
        & {\cellcolor[rgb]{0.85,0.85,0.85}}\textbf{118.96} 
        & MinQ 
        & 1524.30  
        & {\cellcolor[rgb]{0.85,0.85,0.85}}\textbf{159.83} \\

    Mask 25\% 
        & 612.33 
        & {\cellcolor[rgb]{0.85,0.85,0.85}}\textbf{72.57} 
        & Mask 25\% 
        & 1095.14 
        & {\cellcolor[rgb]{0.85,0.85,0.85}}\textbf{75.18} \\

    Mask 50\% 
        & 997.53
        & {\cellcolor[rgb]{0.85,0.85,0.85}}\textbf{86.39} 
        & Mask 50\% 
        & 1095.11
        & {\cellcolor[rgb]{0.85,0.85,0.85}}\textbf{74.52} \\
    \bottomrule
\end{tabular}
\end{table}

Table~\ref{tab:training_complexity} reports training-time complexity under different meta-learning and diffusion settings. All experiments fit on a single RTX 4090 GPU, peak memory remains stable across settings, and wall-clock time grows approximately linearly with the number of cities and diffusion steps. Together, these results suggest that MDRC is practical as an online plug-in defense: its inference cost is substantially lower than prior diffusion-based recovery methods, while its training cost scales predictably with task diversity and diffusion depth.

\begin{table}[ht]
\centering
\caption{Training-Time Complexity under Different Meta-Diffusion Settings}
\label{tab:training_complexity}
   \fontsize{7}{8}\selectfont
\renewcommand{\arraystretch}{1.5}
\setlength{\tabcolsep}{2.6pt}
\begin{tabular}{lcccccccc}
\toprule
Scenario & Tasks & 
\makecell{Diffusion\\Steps} &
\makecell{Meta\\Iter.} &
\makecell{Inner\\Steps} &
\makecell{Wall-clock\\(s)} &
\makecell{Avg/Step\\(s)} &
\makecell{Peak GPU\\(MiB)} \\

\hline
2tasks\_T50   & 2 & 50  & 3 & 30 & 31.93 & 10.64 & 163.06 \\
3tasks\_T50   & 3 & 50  & 3 & 30 & 42.15 & 14.05 & 163.06 \\
3tasks\_T100  & 3 & 100 & 3 & 30 & 42.75 & 14.25 & 163.06  \\
5tasks\_T150  & 5 & 150 & 3 & 30 & 62.57 & 20.86 & 163.06 \\
\bottomrule
\end{tabular}

\end{table}

\takeawaybox{RQ5}{MDRC is practical for online deployment: it cuts denoising and demasking latency by 87.9\% and 91.9\%, respectively, and its training cost scales predictably while remaining feasible on a single RTX 4090 GPU.}

\noindent\textbf{RQ6: Does MDRC improve state recovery fidelity beyond ATT under broader missingness conditions?}

\begin{figure}[ht]
    \centering
    \includegraphics[width=0.8\linewidth]{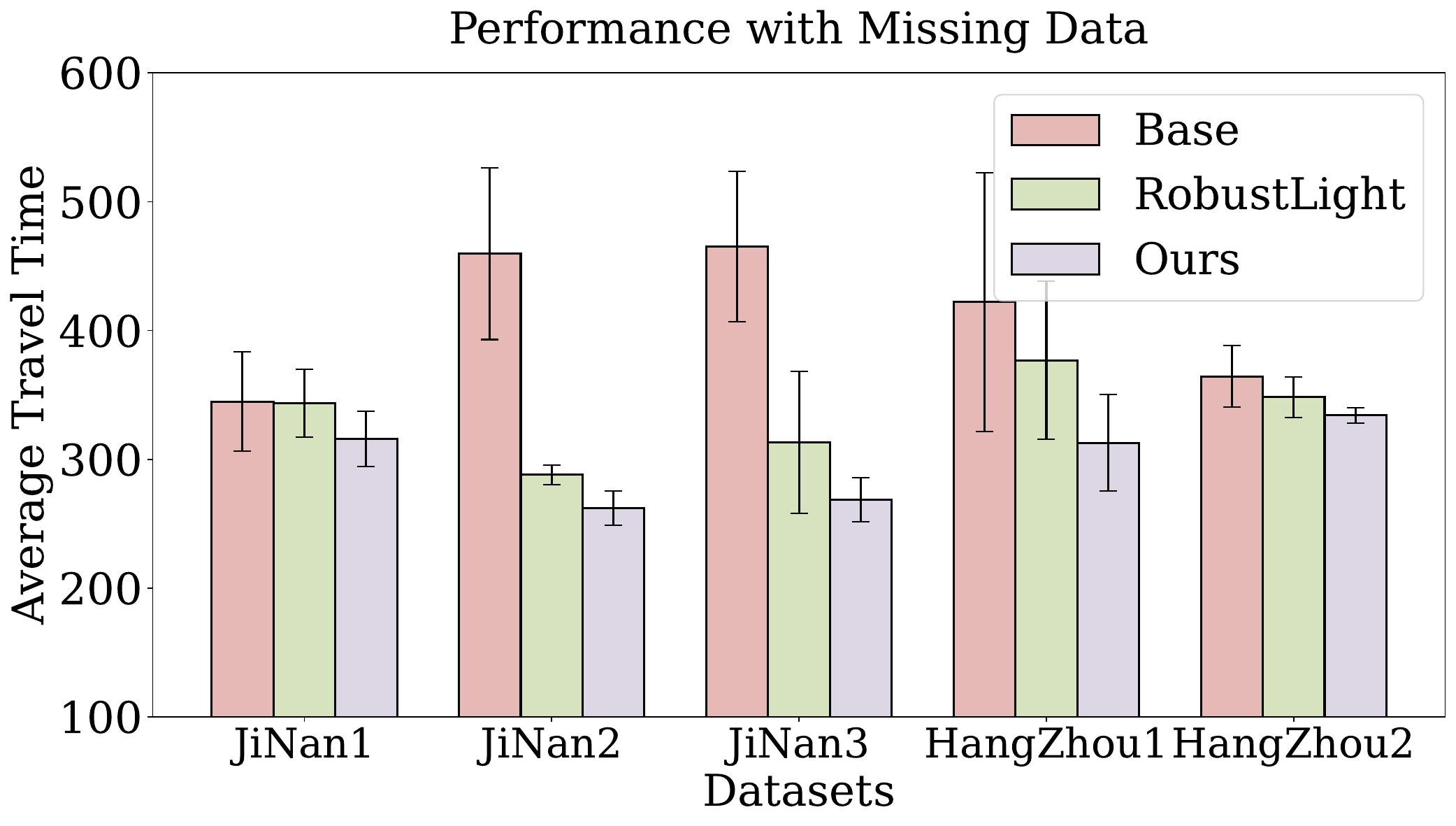}
    \caption{Kriging and random missing.}
    \label{fig:others}
\end{figure}

The previous research questions focus on the primary system-level metric ATT. Here we provide supporting evidence that the ATT gains are tied to better state reconstruction quality rather than policy-specific tuning. Under Kriging Missing (12.5\%, single-intersection-sensor failure) and Random Missing (12.5\%, full-intersection failure), Figure~\ref{fig:others} shows an average improvement of 10.57\%, indicating that MDRC generalizes beyond the main masking configurations studied in RQ2.

Figure~\ref{tab:psnr_and_mae} reports PSNR and MAE on representative datasets and shows that MDRC generally reconstructs cleaner states than RobustLight across both noise and mask settings. Figure~\ref{tab:jinan3_hangzhou1} further evaluates MDRC on top of $\pi$-Light~\cite{pi-light}, again showing consistent improvements under Gaussian and U-rand perturbations. These complementary results are not the primary security claim, but they strengthen the interpretation that MDRC improves downstream control by recovering a better estimate of the true traffic state.

\takeawaybox{RQ6}{The ATT improvements are supported by better state recovery quality. MDRC improves broader settings by 10.57\% on average and generally achieves stronger PSNR and MAE than RobustLight, including when paired with $\pi$-Light.}

\section{Conclusion}
In this paper, we presented MDRC as a post-detection state recovery defense for resilient traffic signal control. Rather than treating robustness as a controller-only learning problem, MDRC protects the sensing-to-control path by reconstructing trustworthy traffic states from corrupted observations and feeding them to the original controller. We provided a principled interpretation of DDIM sampling as a meta-optimization process related to Reptile, which explains why the same recovery model can generalize across cities and adapt with few samples. Experiments showed that MDRC improves performance in unseen cities under stochastic attacks, adaptive policy-targeted attacks, and sensor failures, while reducing denoising and demasking times by 87.9\% and 91.9\%, respectively. MDRC is a practical plug-in defense for scalable real-time urban traffic systems. Future work will explore joint detection-and-recovery pipelines and real-world deployment feedback under multi-intersection coordination.

\section*{Ethics Considerations}

This work studies the resilience of traffic-signal control systems
under corrupted or unavailable sensing and therefore concerns
cyber-physical infrastructure whose malfunction could affect public
safety. We mitigate these risks by conducting adversarial and
sensor-failure experiments in controlled simulation or offline
settings rather than by disrupting operational traffic
infrastructure. Our real-world evaluation uses measured roadside
sensor traces with controlled channel masking, while the
hardware-in-the-loop evaluation preserves the operational safety
constraints enforced by the commercial signal controller. The study
does not involve human subjects, personally identifiable information,
or sensitive user data. The work focuses on post-detection recovery
and does not require compromising live sensing or signal-control
systems. We do not report a previously unknown vulnerability that
requires responsible disclosure. We release sufficient experimental
details and artifacts for reproducibility while anonymizing the
deployment location and infrastructure where appropriate.
\bibliographystyle{IEEEtran}
\bibliography{iclr2026_conference}

\appendix
\section{Open Science}
To enable a thorough and independent evaluation of the contributions presented in this paper, we provide all necessary artifacts required to reproduce and validate our results, including source code, datasets, configuration files, and evaluation scripts.

\textbf{Artifacts provided:}
\begin{itemize}
    \item \textbf{Source code:} Implementation of the proposed method, including training and evaluation pipelines.
    \item \textbf{Datasets:} Processed datasets used in all experiments, along with data preprocessing scripts.
    \item \textbf{Configuration files:} Hyperparameter settings and environment configurations used in all reported experiments.
    \item \textbf{Evaluation scripts:} Scripts used to reproduce all quantitative results and figures reported in the paper.
    \item \textbf{Documentation:} Step-by-step instructions for environment setup, training, and evaluation.
    \item \textbf{Media:} Anonymized HIL runtime videos showing roadside sensing, edge acquisition, server-side phase generation, and commercial-controller feedback.
\end{itemize}

\textbf{Access during double-blind review:} All artifacts are available anonymously at \url{https://anonymous.4open.science/r/MDRC-706C}. Reviewers can directly access the repository using the above link. No authentication or credentials are required. The repository is structured to support end-to-end reproduction of the main experimental results.

\textbf{Reproducibility notes:} We ensure that all reported results can be reproduced using the provided code and datasets under the specified configurations. We also include scripts to facilitate benchmarking and comparison with baselines.

\section{Generative AI Usage}
This work made limited use of an LLM (ChatGPT 5.4) only for language editing, including grammar refinement and typo correction. All conceptual content, technical descriptions, data analysis, and reported results were developed entirely by the authors. No section of the manuscript was generated by generative AI or directly reproduced from LLM outputs. The authors take full responsibility for the content of this paper and the accompanying artifact.

\subsection{Proof of Theorem 1}
\label{detialed_proof}
Recall the deterministic probability flow ODE corresponds to DDIM by \cite{song2020score}. The diffusion step can be written:
\begin{equation}
    \frac{d\bx}{dt} =  \mathbf{f}(\bx, t) - \frac{1}{2}g(t)^2 \nabla \log p_t(\bx).
    \label{eq:true_ode}
\end{equation}
In the DDIM-Reptile framework, the model approximates this field using a learned score network $\mathbf{s}_\theta(\bx, t) \approx \nabla \log p_t(\bx)$. The model's approximate velocity field is:
\begin{equation}
   v(x,t) = \frac{d\bx}{dt} \approx \mathbf{f}(\bx, t) - \frac{1}{2}g(t)^2 \mathbf{s}_\theta(\bx, t).
    \label{eq:model_ode}
\end{equation} 
Let $q_t$ denote the distribution of $x(t)$ under this approximate ODE (with initial distribution $q_T$ at $t=T$). By construction $p_T$ and $q_T$ are the terminal distributions for the true and model processes; if the model precisely matches the chosen noise prior. At time $t=0$, we have $p_0$ (true data) and $q_0$ (model-generated data). Our goal is to bound $W_2(p_0,q_0)$. For each $t\in[0,T]$, let $p_t$ denote the noised data distribution and $q_t$ the corresponding model distribution at time $t$. The score function of $p_t$ is
\begin{equation}
    s^\star(x,t) := \nabla_x \log p_t(x)\,.
\end{equation}
$s_\theta(x,t)$ is the trained score network, and define the score matching loss as
\begin{equation}
    L_{\mathrm{diff}}(\theta)
    \;:=\;
    \frac{1}{2}\int_0^T \lambda(t)\,
      E_{x\sim p_t}\big[\|s^\star(x,t)-s_\theta(x,t)\|^2\big]\,dt,
    \label{eq:Ldiff-def}
\end{equation}
for some positive weighting function $\lambda(t)>0$. According to~\cite{song2020score}, minimizing Eq.~\ref{eq:diff-loss} corresponds to estimating the true score function $s^\star(x,t) = \nabla_x \log p_t(x)$ in the $L^2$ sense. From~\cite{diffusion_warssin} Theorem 1, using Cauchy-Schwarz on that time-integral, we obtain an upper bound in terms of the root of the integrated mean-square error of the score:
\begin{equation}
\begin{aligned}
    W_2(p_0, q_0^{\text{cont}}) \leq & \left[ 2 \int_{0}^{T} g(t)^4 I(t)^2 \, dt \right. \\
    & \cdot \left. \int_{0}^{T} \lambda(t) \mathbb{E}_{p_t}\left[\| \nabla \log p_t(x) - s_\theta(x, t) \|^2\right] \, dt \right]^{1/2} \\
    & + I(T) W_2(p_T, q_T)
\end{aligned}
\label{eq:cont-bound-final}
\end{equation}

The first factor $\int_0^T g(t)^4 I(t)^2 dt$ is a constant determined by the diffusion schedule $g(t)$ and the integrated Lipschitz coefficients $I(t)$. We may therefore write
\begin{equation}
\begin{aligned}
W_2(p_0, q_0^{\mathrm{cont}}) \leq \mathcal{C}_{\mathrm{score}} \sqrt{L_{\mathrm{diff}}(\theta)} + \mathcal{C}_{\mathrm{init}} W_2(p_T, q_T),
\end{aligned}
\end{equation}
for some constant $\mathcal{C}_{\text{score}} = \sqrt{2\int_0^T g(t)^4 I(t)^2dt}$ and $\mathcal{C}_{\text{init}} = I(T)$.

In practice, DDIM approximates the reverse-time dynamics using a
finite sequence of $N$ steps. Let $\Delta t=T/N$ denote the corresponding discretization interval. The resulting update can be viewed as a numerical approximation of the continuous probability-flow
ODE. Under the one-sided Lipschitz assumption below, the discrepancy between the continuous and discretized trajectories scales linearly
with $\Delta t$.
\begin{equation}
x_{t-\Delta t}=x_t+\Delta t\Big(f(x_t,t)-\frac{1}{2} g(t)^2 s_\theta(x_t,t)\Big),
\label{eq:euler_update}
\end{equation}
Equation~\ref{eq:euler_update} is the finite-step numerical approximation used by the DDIM sampler. Here, $\Delta t$ controls only the discretization resolution of the reverse-time dynamics and is distinct from the Reptile outer-loop meta-learning rate $\mu$.
Let $q_0^{\mathrm{cont}}$ denote the distribution induced by the continuous-time ODE, 
and $q_0^{\mathrm{disc}}$ the distribution obtained by the Euler discretization.
Assume the vector field $v(x,t)$ satisfies the one-sided Lipschitz assumption 
proposed by \cite{single-lipts} (Definition 2.1) with coefficient $L_s(t)$:
\begin{equation}
\begin{split}
\langle v(x,t)-v(y,t),\, x-y \rangle 
\;\le\; & L_s(t)\,\|x-y\|^2, \\
\qquad & \forall\, x,y\in\mathbb{R}^d,\; t\in[0,T],
\end{split}
\end{equation}

By Theorem 3.2 and 4.3 of~\cite{single-lipts}, we get the pathwise error bound
\begin{equation}
\sup_{0\le t\le T}
\|x_{\mathrm{cont}}(t)-x_{\mathrm{disc}}(t)\|
\;\le\;
C_{\mathrm{disc}}\Delta t,
\label{eura}
\end{equation}

Let $q_0^{\mathrm{cont}}$ and $q_0^{\mathrm{disc}}$ denote the laws of the continuous and 
Euler-approximated reverse flows at time $t=0$. According to the definition of 2-Wasserstein or Monge-Kantorovich distance~\cite{diffusion_warssin} and combined with Eq.~\ref{eura}, we get:
\begin{equation}
W_2\!\big(q_0^{\mathrm{cont}}, q_0^{\mathrm{disc}}\big)
\;\le\;
\big(\mathbb{E}\|X^{\mathrm{cont}}_0 - X^{\mathrm{disc}}_0\|^2\big)^{1/2}
\;\le\;
C_{\mathrm{disc}}\,\mu,
\label{eq:disc_error}
\end{equation}

From Theorem 4.3 of~\cite{single-lipts}, we get the discretization constant admits the explicit form
\begin{equation}
C_{\mathrm{disc}} = c_0 (C_\tau + C_\chi + 1),
\qquad 
c_0 = \exp\!\Big(\textstyle\int_0^T L_s^+(t)\,dt\Big)\max\{2, B\},
\end{equation}
In particular, the one-sided Lipschitz property prevents exponential error 
amplification in the reverse diffusion ODE, ensuring that discretization error 
accumulates only linearly with $\mu$.

Finally, let $q_0 \equiv q_0^{\mathrm{disc}}$ be the practical model distribution.
By the triangle inequality,
\begin{align}
W_2(p_0, q_0)
&\le
W_2(p_0, q_0^{\mathrm{cont}})
+
W_2(q_0^{\mathrm{cont}}, q_0^{\mathrm{disc}}),
\end{align}
Substituting the continuous-time bound (from the score approximation) 
and the discretization bound \eqref{eq:disc_error}, we obtain
\begin{equation}
W_2(p_0, q_0)
\;\le\;
\mathcal{C}_{\mathrm{score}} \sqrt{L_{\mathrm{diff}}(\theta)}
+
\mathcal{C}_{\mathrm{disc}} \cdot \mu
+
\mathcal{C}_{\mathrm{init}} W_2(p_T, q_T).
\end{equation}
which completes the proof.

\begin{table}[h]
\caption{Hyperparameters}
\label{table: hyperparameter}
\centering
   \fontsize{7}{8.5}\selectfont
\renewcommand{\arraystretch}{1.2}
\setlength{\tabcolsep}{2pt}
\begin{tabular}{ccc} 
\toprule
\textbf{Hyperparameter Type} & \textbf{Hyperparameter} & \textbf{Setting} \\ 
\midrule

\multirow{4}{*}{\begin{tabular}[c]{@{}c@{}}UNet/Model\\Hyperparameter\end{tabular}} 
    & embed\_dim           & 64     \\
    & state\_dim           & 20/32  \\
    & action\_dim          & 4      \\
    & hidden\_dim          & 256 \\
\midrule

\multirow{13}{*}{\begin{tabular}[c]{@{}c@{}}Diffusion Training\\Hyperparameter\end{tabular}} 
    & non\_markovian\_step  & 6                           \\
    & condition\_length     & 4                           \\
    & beta schedule         & \multicolumn{1}{l}{3.0651, 24.552, -3.1702}               \\
    & discount($\gamma$)    & 0.99                        \\
    & target critic($\tau$) & 0.005                       \\
    & diffusion timestep    & 100                         \\
    & batch size            & 16                          \\
    & buffer capacity       & 240                         \\
    & optimizer             & Adam                        \\
    & learning rate         & 0.0003                      \\
    & meta learning rate $\mu$    & 0.1                         \\
    & epochs(meta/single)   & 25 / 90                     \\
    & hidden size           & 256                         \\
    & attention embed\_dim  & 64                          \\
\midrule

\multirow{12}{*}{\begin{tabular}[c]{@{}c@{}}TSC RL Agent\\Training Hyperparameter\end{tabular}} 
    & discount($\gamma$)     & 0.8                  \\
    & buffer capacity        & 12000                \\
    & epochs                 & 80                   \\
    & batch\_size            & 20                   \\
    & learning\_rate         & 0.001                \\
    & target update time     & 5                    \\
    & normal factor          & 20                   \\
    & loss function          & mean\_squared\_error \\
    & optimizer              & Adam                 \\
    & learning rate          & 0.001                \\
    & patience               & 10                   \\
    & epsilon (init/min/decay) & 0.8 / 0.2 / 0.95   \\
    & D\_DENSE               & 20                   \\

\bottomrule
\end{tabular}
\end{table}

\subsection{Hyperparameters}
\label{hyperparameter}
The Table~\Cref{table: hyperparameter} outlines the hyperparameters used for training the proposed model, categorized into UNet/Model Hyperparameters, Diffusion Training Hyperparameters, and TSC RL Agent Training Hyperparameters. Key settings include an embedding dimension of 64 and a hidden dimension of 256 for the UNet model, with state and action dimensions tailored to 20/32 and 4, respectively. Diffusion training employs a non-Markovian step of 6, a beta schedule of [3.0651, 24.552, -3.1702], a diffusion timestep of 100, and an Adam optimizer with a learning rate of 0.0003, alongside meta and single epoch settings of 25 and 90. The TSC RL agent is configured with a discount factor of 0.8, a buffer capacity of 12,000, a batch size of 20, and an Adam optimizer with a learning rate of 0.001, incorporating an epsilon greedy strategy with initial, minimum, and decay values of 0.8, 0.2, and 0.95, respectively.


\subsection{Sensitivity to Imperfect Fault Detection}
\label{app:detection-sensitivity}

MDRC assumes that an upstream sensor-health or anomaly-detection
module identifies unreliable observations before recovery. To evaluate
its sensitivity to imperfect triggering, we impose 50\% sensor
missingness and vary the detector recall over the truly missing
channels. Detector precision is assumed to be perfect, while recall is
set to approximately 60\% or 80\%. MDRC reconstructs only the channels
identified as missing; undetected missing channels remain zero-filled
in the controller input. Accordingly, the reference condition is the
same 50\%-missing observation without recovery rather than a
clean-sensing baseline.

We evaluate the resulting detector--MDRC pipeline using 3,600-second
closed-loop CityFlow rollouts. Each configuration is evaluated over
three paired trials, and results are reported as mean $\pm$ standard
deviation. For Jinan, recovery uses the source-city model. For
Hangzhou, we evaluate both zero-shot transfer from Jinan and few-shot
adaptation using target-city calibration data. Average travel time
(ATT) and mean waiting vehicles are minimized, whereas throughput is
maximized. Percentage changes are calculated relative to the
corresponding 50\%-missing, no-recovery baseline.
\begin{table*}[t]
\centering
\caption{Closed-loop performance under 50\% sensor missingness
with imperfect missing-channel detection. Results are reported as
mean $\pm$ standard deviation over three paired trials. Percentages
denote changes relative to the no-recovery baseline of the
corresponding network.}
\label{tab:detection-rate-recovery}

\footnotesize
\setlength{\tabcolsep}{4.5pt}
\renewcommand{\arraystretch}{0.8}

\begin{tabular}{@{}llcccc@{}}
    \toprule
    \textbf{Network}
    & \textbf{Recovery model}
    & \textbf{Detection recall}
    & \textbf{ATT (s) $\downarrow$}
    & \textbf{Throughput (veh/h) $\uparrow$}
    & \textbf{Waiting $\downarrow$} \\
    \midrule

    \multirow{3}{*}{\textbf{$JiNan_1$}}
    & None (baseline)
    & --
    & $704.31 \pm 200.27$
    & $3414 \pm 760$
    & $743.27 \pm 339.02$ \\

    & Source model
    & 60\%
    & \makecell{$484.58 \pm 90.21$\\[-1pt]
      \scriptsize $(-31.2\%)$}
    & \makecell{$4281 \pm 356$\\[-1pt]
      \scriptsize $(+25.4\%)$}
    & \makecell{$358.43 \pm 108.60$\\[-1pt]
      \scriptsize $(-51.8\%)$} \\

    & Source model
    & 80\%
    & {\cellcolor[rgb]{0.85,0.85,0.85}
      \makecell{\textbf{$386.30 \pm 20.53$}\\[-1pt]
      \scriptsize\textbf{$(-45.2\%)$}}}
    & {\cellcolor[rgb]{0.85,0.85,0.85}
      \makecell{\textbf{$4704 \pm 67$}\\[-1pt]
      \scriptsize\textbf{$(+37.8\%)$}}}
    & {\cellcolor[rgb]{0.85,0.85,0.85}
      \makecell{\textbf{$233.26 \pm 29.92$}\\[-1pt]
      \scriptsize\textbf{$(-68.6\%)$}}} \\

    \midrule

    \multirow{5}{*}{\textbf{$HangZhou_1$}}
    & None (baseline)
    & --
    & $667.00 \pm 54.85$
    & $3089 \pm 293$
    & $597.71 \pm 92.20$ \\

    & Zero-shot
    & 60\%
    & \makecell{$601.83 \pm 56.65$\\[-1pt]
      \scriptsize $(-9.8\%)$}
    & \makecell{$3461 \pm 361$\\[-1pt]
      \scriptsize $(+12.1\%)$}
    & \makecell{$416.09 \pm 135.51$\\[-1pt]
      \scriptsize $(-30.4\%)$} \\

    & Zero-shot
    & 80\%
    & \makecell{$546.64 \pm 43.37$\\[-1pt]
      \scriptsize $(-18.0\%)$}
    & \makecell{$3783 \pm 271$\\[-1pt]
      \scriptsize $(+22.5\%)$}
    & \makecell{$315.99 \pm 87.96$\\[-1pt]
      \scriptsize $(-47.1\%)$} \\

    \cmidrule(lr){2-6}

    & Few-shot
    & 60\%
    & \makecell{$585.18 \pm 37.72$\\[-1pt]
      \scriptsize $(-12.3\%)$}
    & \makecell{$3518 \pm 283$\\[-1pt]
      \scriptsize $(+13.9\%)$}
    & \makecell{$388.55 \pm 105.25$\\[-1pt]
      \scriptsize $(-35.0\%)$} \\

    & Few-shot
    & 80\%
    & {\cellcolor[rgb]{0.85,0.85,0.85}
      \makecell{\textbf{$535.84 \pm 32.44$}\\[-1pt]
      \scriptsize\textbf{$(-19.7\%)$}}}
    & {\cellcolor[rgb]{0.85,0.85,0.85}
      \makecell{\textbf{$3801 \pm 244$}\\[-1pt]
      \scriptsize\textbf{$(+23.1\%)$}}}
    & {\cellcolor[rgb]{0.85,0.85,0.85}
      \makecell{\textbf{$296.76 \pm 82.77$}\\[-1pt]
      \scriptsize\textbf{$(-50.4\%)$}}} \\

    \bottomrule
\end{tabular}

\vspace{2pt}
\parbox{0.90\textwidth}{\scriptsize
Detection rates denote recall over truly missing channels, with
detector precision fixed at 100\%. Undetected missing channels
remain zero-filled and are not reconstructed. Gray cells denote
the best recovery result for each network. Lower ATT and waiting
are better, whereas higher throughput is better.}

\end{table*}
\paragraph{Results.}
MDRC improves all three control metrics at both detector operating
points. In Jinan, 80\% detection recall reduces ATT by 45.2\%,
increases throughput by 37.8\%, and reduces the mean waiting queue by
68.6\% relative to the unrecovered baseline. Even at 60\% recall,
MDRC retains substantial improvements of 31.2\%, 25.4\%, and 51.8\%,
respectively.

The same pattern persists after transfer to Hangzhou. With zero-shot
recovery, 80\% recall improves ATT, throughput, and waiting by 18.0\%,
22.5\%, and 47.1\%, while 60\% recall yields improvements of 9.8\%,
12.1\%, and 30.4\%. Few-shot adaptation further improves performance,
reaching reductions of 19.7\% in ATT and 50.4\% in waiting, together
with a 23.1\% throughput increase at 80\% recall.

Across both networks and recovery models, the 80\% operating point
consistently outperforms the 60\% setting. Nevertheless, MDRC remains
beneficial when 40\% of the missing channels are not detected. These
results show that recovery does not require perfect detector recall,
although its traffic-level benefit increases as more faulty channels
are correctly identified. The comparison is made against severe
50\% sensor missingness and should not be interpreted as equivalence
between recovered and clean observations.
\subsection{Additional Experiments}

\begin{figure}[ht]
    \centering
    \includegraphics[width=0.8\linewidth]{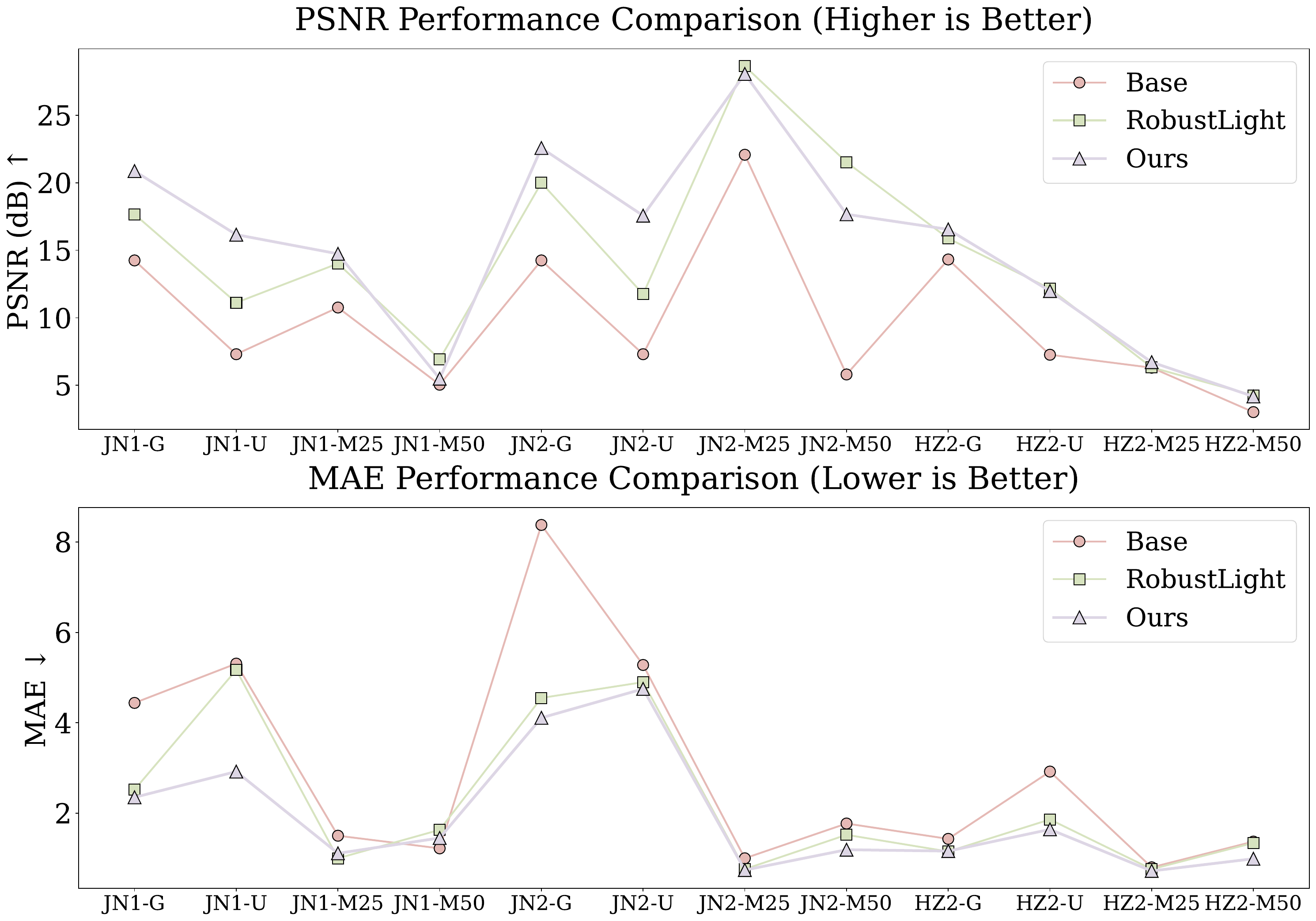}
    \caption{PSNR and MAE comparison with RobustLight.}
    \label{tab:psnr_and_mae}
\end{figure}

\begin{figure}[ht]
    \centering
    \includegraphics[width=1\linewidth]{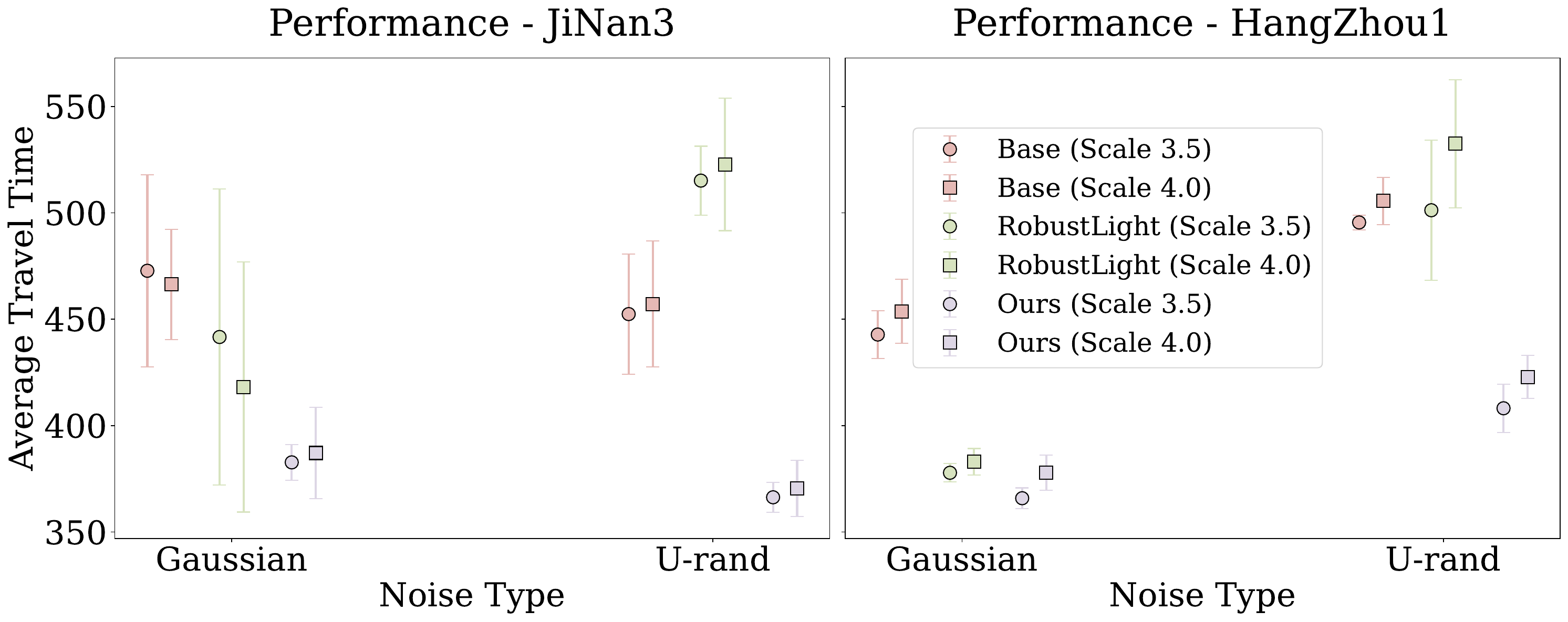}
    \caption{Performance on $JiNan_3$ and $HangZhou_1$ under different noise types and scales based on $\pi$-Light.}
    \label{tab:jinan3_hangzhou1}
\end{figure}
\begin{table}[ht]
\caption{Performance comparison on New York-scale.}
\label{table:newyork_scale}
\centering
\fontsize{7}{8}\selectfont
\renewcommand{\arraystretch}{1.15}
\setlength{\tabcolsep}{3.5pt}

\begin{tabular}{lccc}
    \toprule
    Corruption & Noisy-base & RobustLight & \textbf{Our} \\
    \midrule

    \multicolumn{4}{c}{\textbf{NewYork1}} \\
    \midrule
    Gaussian-3.5
        & 1279.46
        & 1168.11
        & {\cellcolor[rgb]{0.85,0.85,0.85}}\textbf{1154.30} \\

    U-rand-3.5
        & 1527.23
        & 1359.38
        & {\cellcolor[rgb]{0.85,0.85,0.85}}\textbf{1330.24} \\

    Mask-25\%
        & 1119.34
        & {\cellcolor[rgb]{0.85,0.85,0.85}}\textbf{1098.55}
        & 1130.83 \\

    \midrule
    \multicolumn{4}{c}{\textbf{NewYork2}} \\
    \midrule
    Gaussian-3.5
        & 1653.00
        & 1501.80
        & {\cellcolor[rgb]{0.85,0.85,0.85}}\textbf{1427.23} \\

    U-rand-3.5
        & 1788.00
        & 1697.33
        & {\cellcolor[rgb]{0.85,0.85,0.85}}\textbf{1545.65} \\

    Mask-25\%
        & 1321.09
        & 1376.90
        & {\cellcolor[rgb]{0.85,0.85,0.85}}\textbf{1286.63} \\

    \bottomrule
\end{tabular}
\end{table}
We tested NewYork with few-shot transfer, where MDRC achieves 1091.32 ATT, better than the noisy-baseline. The zero-shot NewYork Mask case is failure maybe the downstream controller is graph-centralized. If several intersection states are not recovered well, local errors propagate through the graph controller and amplify congestion. This differs from decentralized controllers such as MaxPressure. Our detector sensitivity analysis varies recall while holding precision at 100\%; therefore it evaluates missed detections but not false-positive recovery triggers.

\end{document}